\documentclass[usenatbib,usegraphicx,letterpaper]{mn2e}
\usepackage[totalwidth=480pt,totalheight=680pt]{geometry}

\usepackage{amssymb}
\usepackage{epsfig}
\usepackage{amsmath}
\usepackage[usenames]{color}
\usepackage[draft]{hyperref}
\definecolor{purple}{RGB}{160,32,240}
\hypersetup{colorlinks=true,citecolor=blue,linkcolor=purple,filecolor=black,runcolor=black}

\newcommand{\dd}{\mathrm{d}}

\newcommand{\beq}{\begin{equation}}
\newcommand{\eeq}{\end{equation}}
\newcommand{\beqray}{\begin{eqnarray}}
\newcommand{\eeqray}{\end{eqnarray}}

\newcommand{\ben}{\begin{enumerate}}
\newcommand{\een}{\end{enumerate}}
\newcommand{\bit}{\begin{itemize}}
\newcommand{\eit}{\end{itemize}}

\newcommand{\mvir}{M_{\mathrm{vir}}}
\newcommand{\rvir}{R_{\mathrm{vir}}}
\newcommand{\msun}{M_{\odot}}

\newcommand{\rhill}{R_{\rm Hill}}
\newcommand{\rhillmin}{R_{\rm Hill-min}}

\newcommand{\mar}{\dd\mvir/{\rm dt}}

\newcommand{\Mprim}{M_{\rm  prim}}
\newcommand{\Msec}{M_{\rm  sec}}

\newcommand{\dphys}{D}

\newcommand{\taudyn}{\tau_{\rm dyn}}

\newcommand{\unit}[1]{\mathrm{#1}}

\newcommand{\mpc}{\unit{Mpc}}

\usepackage{epsfig}  \usepackage{graphicx}   \usepackage{rotating}

\newcommand{\aap}{A\&A~}

\newcommand{\apjl}{ApJL~}
\newcommand{\apjs}{ApJS~}

\begin{document}

\title[On the Origin of Conformity]
{On the Physical Origin of Galactic Conformity}

\author[Hearin, Behroozi \& van den Bosch]
{Andrew~P.~Hearin$^{1}$, Peter S. Behroozi$^{2}$, \&  Frank~C.~van den Bosch$^{3}$\\
$^1$Yale Center for Astronomy \& Astrophysics, Yale University, New Haven, CT\\
$^2$Space Telescope Science Institute, Baltimore, MD 21218, USA \\
$^3$Department of Astronomy, Yale University, P.O. Box 208101, New Haven, CT}

\maketitle

\begin{abstract}
Correlations between the star formation rates (SFRs) of nearby galaxies (so-called {\em galactic conformity}) have been observed for projected separations up to 4 Mpc, an effect not predicted by current semi-analytic models. We investigate correlations between the mass accretion rates ($\mar$) of nearby halos as a potential physical origin for this effect. We find that pairs of host halos ``know about'' each others' assembly histories {\em even when their present-day separation is greater than thirty times the virial radius of either halo.} These distances are far too large for direct interaction between the halos to explain the correlation in their $\mar.$  Instead, halo pairs at these distances reside in the same large-scale tidal environment, which regulates $\mar$ for both halos. Larger halos are less affected by external forces, which naturally gives rise to a mass dependence of the halo conformity signal. SDSS measurements of galactic conformity exhibit a qualitatively similar dependence on stellar mass, including how the signal varies with distance. Based on the expectation that halo accretion and galaxy SFR are correlated, we predict the scale-, mass-  and redshift-dependence of large-scale galactic conformity, finding that the signal should drop to undetectable levels by $z\gtrsim1$.  These predictions are testable with current surveys to $z\sim 1$; confirmation would establish a strong correlation between dark matter halo accretion rate and central galaxy SFR.
\end{abstract}



\section{Introduction}
\label{sec:intro}

In the $\Lambda$CDM paradigm, galaxies form at the centers of collapsed, gravitationally self-bound halos of dark matter.  Across most of cosmic time, galaxy stellar mass exhibits a tight statistical connection with dark matter halo mass \citep[e.g.,][]{conroy_wechsler09,moster10,moster13,behroozi10,behroozi13b,behroozi13,Firmani10,leauthaud11a,leitner12,yang12,Bethermin13,Wang12,Mutch13,Lu14}.  Yet, \textit{individual} galaxies enjoy a diversity of star formation rates \citep[e.g.,][]{Brinchmann04,Noeske07,Salim07,Brammer11,Moustakas13,Muzzin13}, which has proven more difficult to physically and self-consistently link to halo properties \citep{YuLu14b,Genel14,Behroozi15}.

A recent insight into this link is {\em galactic conformity}: the tendency of neighboring galaxies to have similar specific star formation rates (SSFRs), colors, gas fractions, and morphologies.  This effect was first observed for satellites of larger (i.e., ``central'') galaxies \citep{weinmann06b,Wang10,Kauffmann10,Wang12c,Robotham13,Phillips14,Knobel15}, but later also found to occur for galaxies separated by up to 4 Mpc in projection \citep{kauffmann_etal13}.  Naively, this larger scale ``2-halo conformity" \citep[see][]{hearin_etal14} would be difficult to explain via internal baryonic processes alone, as these correlations extend over 1000 times the half-light radii of the galaxies and over 10 times the virial radii of the halos concerned.\footnote{Observations of ``1-halo conformity" also remain to be successfully explained, though this paper will chiefly be concerned with correlations beyond the halo's virial radius. See, however, \S\ref{subsec:onehaloconformity}.} Two-halo conformity is evidently quite challenging to correctly predict: the signal in contemporary semi-analytic models \citep{guo_etal11} and the state-of-the-art Illustris hydrodynamical simulation (Bray et al., private communication) is nowhere near as strong as in SDSS observations. 

Although two-halo conformity in galaxy SSFR may seem puzzling, in fact many properties of dark matter halos are correlated on these scales \citep{hahn_etal07b,hahn_etal07,Wang11,Shi15}. Such correlations are not at all puzzling, as nearby halos form out of the same over-dense patch in the cosmic density field. This phenomenon has been investigated extensively in the context of ``assembly bias''---a dependence of the clustering of halos on properties besides mass \citep{gao_etal05,Wechsler06,wetzel_etal07,Wang07,Keselman07,gao_white07,croton_etal07,wu08,li_etal08,dalal_etal08,faltenbacher_white10,lacerna11,lacerna12,wang_etal13,zentner_etal13}.  Previous studies have also established long-range correlations between tidal forces and internal halo properties and/or halo assembly history \citep{Wang07,hahn_etal07b,hahn_etal07,hahn_etal09,Wang11,Behroozi14,Shi15}. There is no shortage of halo properties with large-scale correlations. From this perspective, two-halo conformity---especially conformity between the SSFRs of central galaxies---should naturally arise if star-formation history were coupled sufficiently strongly to one or more halo properties that exhibit large-scale correlations. 

The dark matter accretion rate of a halo is tightly coupled with the accretion rate of baryonic gas \citep[see][for a recent demonstration]{wetzel_nagai14}. This tight coupling provides a theoretical basis for connecting the SFR of a galaxy to its halo's dark matter accretion rate ($\mar$). In this paper, we investigate whether a connection of this type could give rise to an SDSS-like conformity signal, and predict how parameters of the conformity measurement (distance, mass, and redshift) affect the strength of the signal.

This paper is organized as follows. We describe the N-body simulation we use in \S \ref{sec:sim}, and in \S \ref{sec:results} we demonstrate that halo accretion rates are correlated over the same range of scales as SDSS galaxy star-formation rates.  We identify the root cause of this phenomenon as correlations in the tidal forces experienced by nearby halos.  This results in a unique prediction for the halo mass and redshift-dependence of two-halo conformity under our hypothesis that galaxy star formation rate is coupled to halo accretion rate.  We discuss tests of these predictions and implications in \S \ref{sec:discussion}, and summarize our primary conclusions in \S \ref{sec:conclusions}.  

\section{Simulation \& Halos}
\label{sec:sim}

We use the \textit{Bolshoi} simulation \citep{Bolshoi} for all our analysis.  \textit{Bolshoi} follows 2048$^3$ particles ($\sim$ 8 billion) in a cubic box of side length 250 $h^{-1}$ Mpc using the \textsc{art} code \citep{kravtsov_etal:97}.  Its mass resolution ($1.36\times 10^8$ $h^{-1} \msun$) and force resolution (1 $h^{-1}$ kpc) allow it to resolve halos down to $\sim10^{10}\msun.$  \textit{Bolshoi} adopts a flat $\Lambda$CDM cosmology ($\Omega_M=0.27$, $\Omega_{\Lambda} = 0.73$, $h=0.7$, $\sigma_8 = 0.82$, $n_s = 0.95$) which is very similar to the WMAP9 best-fit cosmology \citep{WMAP9}.  Halos were found using the \textsc{Rockstar} halo finder \citep{Rockstar}, which associates particles to peaks in the phase-space density; the finder also makes use of temporal information to improve stability in major mergers. \textsc{Rockstar} has been shown to have excellent recovery of both halo and subhalo properties \citep{Knebe11,Onions12}.  Merger trees were constructed using the \textsc{Consistent Trees} algorithm \citep{BehrooziTree}, which predicts the gravitational evolution of halos from one snapshot to another so as to identify and repair inconsistencies \citep[see also][]{Srisawat13}. 

Throughout the paper, we neglect subhalos and exclusively use host halos in all of our calculations. We will quote results for a halo's mass accretion rate, $\mar.$ For a halo identified at cosmic time $t_1,$ we compute this quantity as 
\beq
\label{eq:taudyndef}
\mar \equiv \frac{\mvir(t_1-\taudyn(t_1)) - \mvir(t_1)}{\taudyn(t_1)};
\eeq
in Eq.~\ref{eq:taudyndef}, $\mvir(t_1-\taudyn(t_1))$ is the virial mass of the halo's main progenitor at time $t_1-\taudyn(t_1)$ as identified by \textsc{Rockstar}, 
where $\taudyn = (G\Delta_{\rm vir}\rho_{\rm crit})^{-1/2},$ with $\Delta_{\rm vir}$ being the virial overdensity \citep{bryan_norman98}.

\section{Results}
\label{sec:results}

In  \S\ref{subsec:tides}, we review the physics linking halo accretion rates to large-scale environment. We then present the primary result in \S\ref{subsec:upshot}: the mass accretion rates of pairs of halos are correlated for pair separations out to $10$ $\mpc,$ a phenomenon we dub {\em halo accretion conformity.} In \S\ref{subsec:upshot} we also demonstrate that the underlying physics of this effect is quite intuitive, and is due to the halo pair mutually evolving in the same large-scale tidal environment. We show the redshift-dependence of halo accretion conformity in  \S\ref{subsec:highzresults}. 

\subsection{Tidal Forces and Mass Accretion Rates}
\label{subsec:tides}

For a detailed review of the connection between the mass accretion rate of a halo and its large-scale environment, we refer the reader to \citet{hahn_etal09}.  Here, we summarize the basic physical picture.

\begin{figure}
\centering
\includegraphics[width=7.5cm]{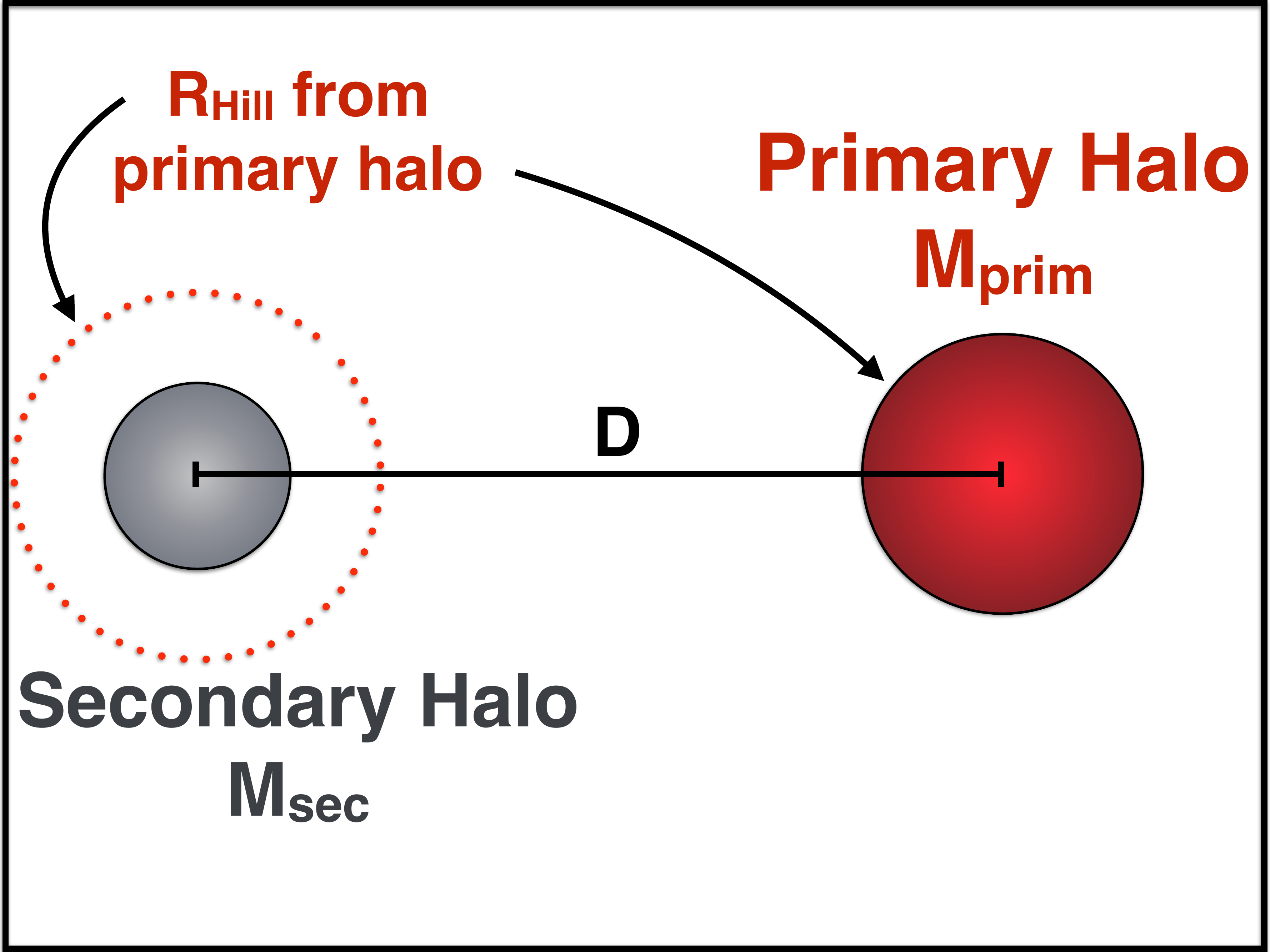}
\caption{ {\bf Diagram of the Hill Radius $\rhill.$} Due to the tidal field produced by the primary, dark matter particles beyond $\rhill=\dphys\left(\Msec/3\Mprim\right)^{1/3}$ cannot accrete onto and remain bound to the secondary halo. Thus halos in {\em stronger} tidal fields have {\em smaller} Hill Radii. When $\rhill$ is roughly equal to $\rvir,$ the virial radius of the secondary, we should expect the secondary halo to experience a significant suppression of its mass accretion rate $\mar.$
}
\label{fig:cartoon}
\end{figure}

Consider a pair of halos separated by distance $\dphys$, calling the larger halo the \textit{primary} and the smaller halo the \textit{secondary}; i.e., so that $\Mprim>\Msec$. Because of the tidal force exerted by the primary upon the secondary, the only stable circular orbits around the secondary halo are for particles whose distance is smaller than the Hill Radius, 
\begin{eqnarray}
\rhill & = & \dphys\left(\Msec/3\Mprim\right)^{1/3} \nonumber\\
 & = & R_\mathrm{sec} (D / \sqrt[3]{3} R_\mathrm{prim})
\label{eq:rhilldef}
\end{eqnarray}
The {\em smaller} the distance between the halos, the {\em stronger} is the tidal force, and the {\em smaller} is the radius around the secondary that circular orbits are stable (i.e., the smaller is $\rhill$).  The Hill Radius is thus an upper bound on the spatial extent of newly infalling material that can remain gravitationally bound to the secondary halo. In this way, the presence of the primary regulates the mass accretion rate of the secondary.  Figure~\ref{fig:cartoon} shows a cartoon illustration of the Hill Radius for a pair of primary and secondary halos. 

For a secondary halo of virial radius $\rvir,$ when {\em any} primary halo exerts a tidal force resulting in $\rhill\sim\rvir,$ we should expect the secondary halo to experience a strong tidal field that inhibits its mass accretion rate. Motivated by this natural expectation, we quantify the strength of a halo's large-scale tidal field with $\rhillmin,$ defined as the minimum $\rhill$ due to any of the halo's more massive neighbors $M>\Msec,$ 
\beq
\rhillmin \equiv {\rm MIN}\left\{\rhill\right\}_{M>\Msec}
\label{eq:rhillmin}
\eeq

In Figure~\ref{fig:mar_vs_rhill}, we show that the  above natural expectation is borne out quantitatively. For $z=0$ {\em Bolshoi} host halos in a narrow range of virial mass\footnote{All mass bins in this paper are $0.35$ dex in total width.} $\mvir\approx10^{11}\msun,$ we plot the mean present-day mass accretion rate $\mar$ of secondary halos as a function of $\rhillmin/\rvir.$ The gray band in Figure~\ref{fig:mar_vs_rhill} shows the error-on-the-mean, which we estimate here and throughout the paper via jackknifing 125 sub-volumes of the simulated box. The blue shaded vertical region shows the inner $50\%$ of the $\rhillmin-$distribution of $10^{11}\msun$ halos.

Figure~\ref{fig:mar_vs_rhill} illustrates the following rough rule of thumb. For secondary halos with $\rhillmin\gtrsim3\rvir,$ there is only a very weak dependence of $\mar$ on environment. On the other hand, when $\rhillmin\lesssim3\rvir$ we can see that $\mar$ drops precipitously. On average, $z=0$ secondary halos with $\rhillmin\lesssim\rvir$ have accreted no mass over the past dynamical time $\taudyn\approx2$Gyr. 

As shown in \citet{hahn_etal09}, there is a pronounced halo mass dependence to the trend shown in Fig.~\ref{fig:mar_vs_rhill}. For halos less massive than the collapse mass ($M_{\rm coll}$, which is $\approx 10^{12.6}\msun$ at $z=0$), the mass accretion rate exhibits the sharp $\rhillmin-$dependence shown in Fig.~\ref{fig:mar_vs_rhill}; for more massive halos ($\mvir>M_{\rm coll}$), there is only a very weak dependence. Relative to higher-mass halos, the accretion rates $\mar$ of lower-mass halos are more strongly affected by their large-scale environment. We will see in the next section how this previously-established trend  manifests in halo accretion conformity.

\subsection{Halo Accretion Conformity}
\label{subsec:upshot}

\subsubsection{Definition \& observational motivation}

As discussed in \S\ref{sec:intro}, recent SDSS measurements \citep{kauffmann_etal13} have shown that the SSFR of a sample of ``primary'' central galaxies is correlated with the mean SSFR of all the ``secondary'' galaxies in the neighborhood of the primaries; the correlation persists out to projected separations of $R_{\rm p}\approx4$ $\mpc.$ Furthermore, the strength of this correlation weakens as the stellar mass of the primaries increases. 

In this section, we show that dark matter halos exhibit directly analogous trends in their mass accretion rates. Primary dark matter halos with above-average $\mar$ tend to be located in an environment populated with secondary halos that are also rapidly accreting mass. The converse is also true for primary halos with below-average $\mar.$  We refer to this phenomenon as {\em halo accretion conformity.} 

\subsubsection{Basic physical picture}
\label{subsubsec:intuition}

\begin{figure}
\centering
\includegraphics[width=9cm]{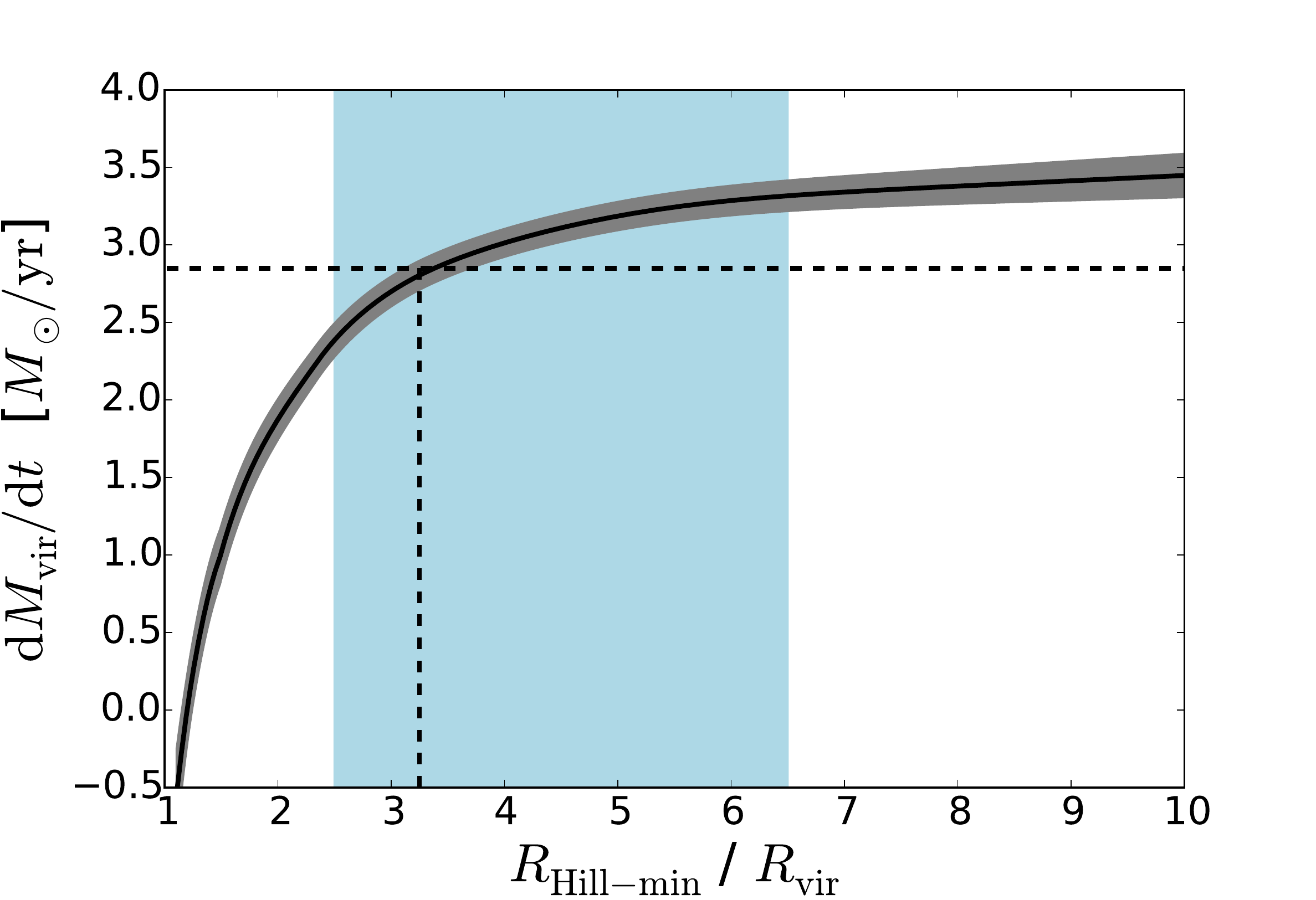}
\caption{ {\bf Strong tidal fields stifle mass accretion rates.} For each secondary halo with $\mvir=10^{11}\msun,$ we compute $\rhillmin,$ the minimum Hill Radius due to any more massive halo in the simulation. The black curve shows the present-day mass accretion rate $\mar$ of the secondary halos as a function of $\rhillmin;$ the gray band shows our jackknife estimation of the error-on-the-mean. The shaded blue region shows the inner $50\%$ of the $\rhillmin-$distribution of $10^{11}\msun$ halos. The horizontal dashed line shows the median $\dd\mvir/{\rm dt}$ of all $10^{11}\msun$ halos; the vertical dashed line guides the eye to the corresponding horizontal axis value. Tidal forces with $R_{\rm hill-min} \lesssim 3\rvir$ dramatically stifle accretion onto the secondary; for $\rhillmin>3\rvir,$ the large-scale environment does not significantly impact $\mar.$ 
}
\label{fig:mar_vs_rhill}
\end{figure}


\begin{figure*}
\centering
\includegraphics[width=13cm]{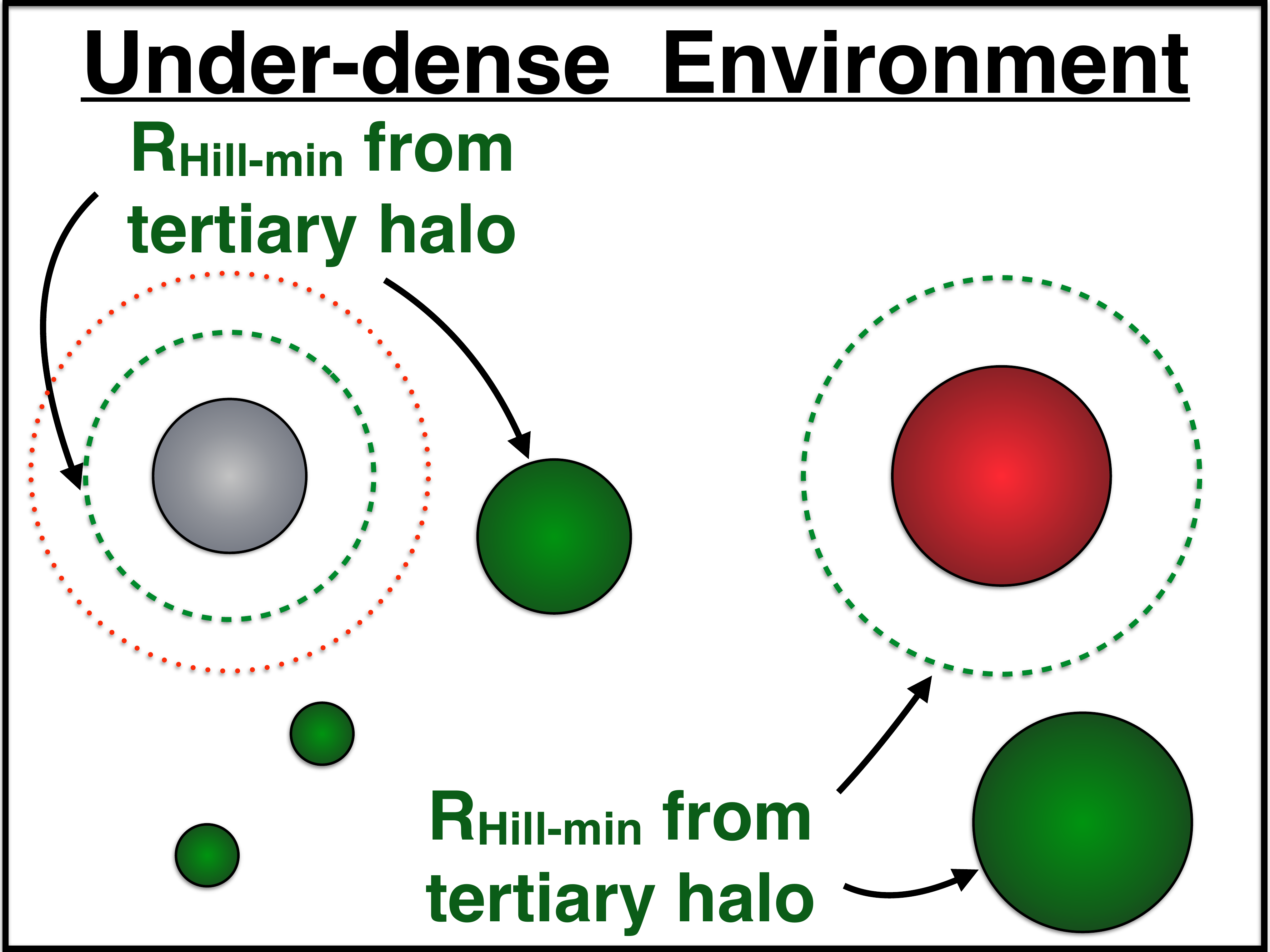}
\includegraphics[width=13cm]{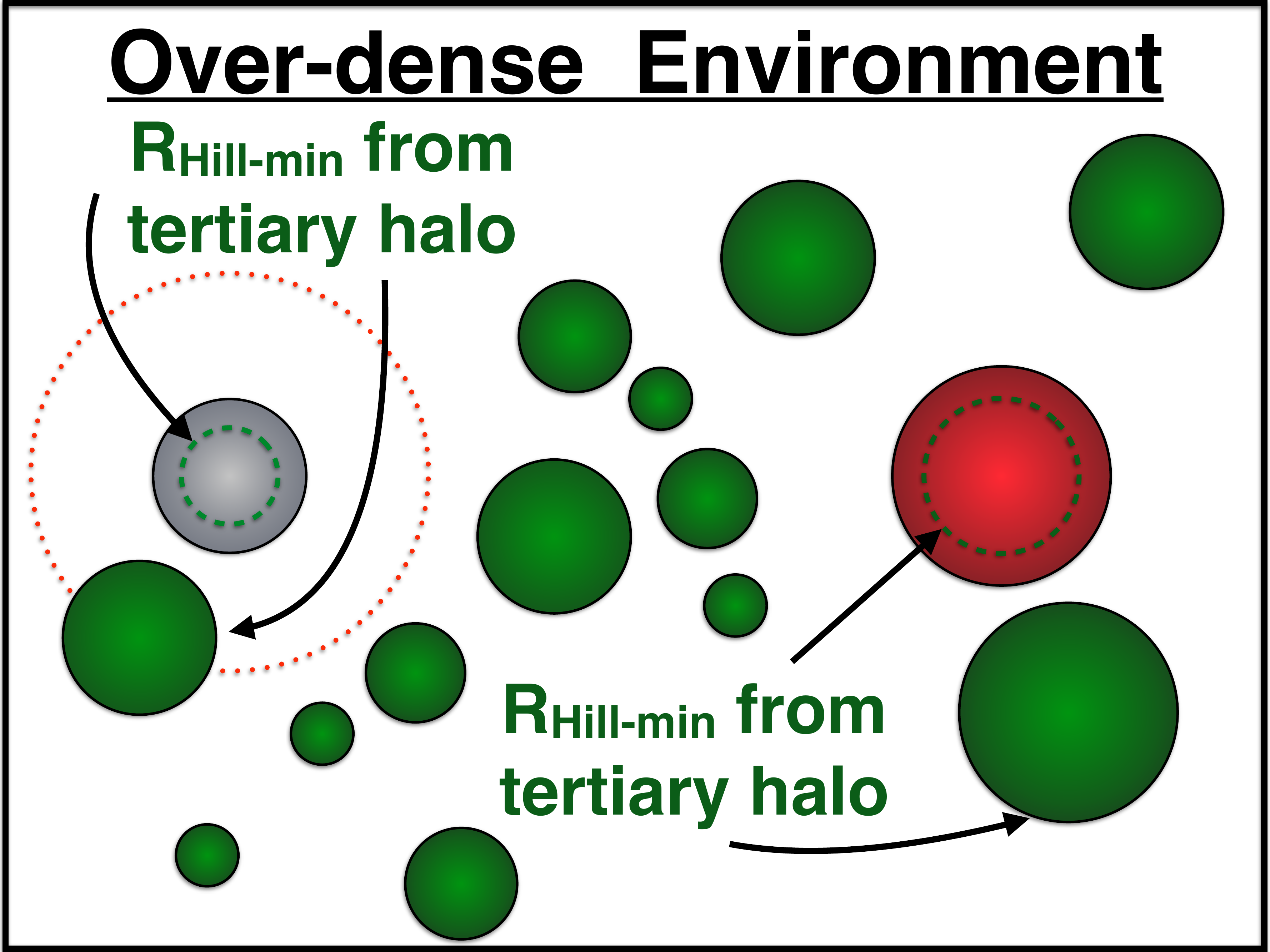}
\caption{ {\bf The physics of halo accretion conformity.} The primary (red) and secondary (gray) halos are so distant that the direct tidal force between them has a negligible influence on either of their mass accretion rates. However, their accretion rates can still be correlated because the halos mutually evolve in the same large-scale environment. A halo evolving in a crowded field of other halos will tend to be located near another massive tertiary halo; this results in a strong tidal field, a small $\rhill$, and a suppressed mass accretion rate.  Conversely, halos evolving in a void-like environment tend to be more isolated, with fewer massive tertiary halos available to exert a significant tidal force.  Since this phenomenon pertains to both primaries and secondaries alike, there is a natural tendency for correlated assembly histories to arise between halos evolving in a similar tidal environment; this tendency is the physical origin of halo accretion conformity. 
}
\label{fig:cartoon2}
\end{figure*}


Before presenting our measurements of conformity in simulated halos, let's first consider how the physics discussed in \S\ref{subsec:tides} should naturally lead to halo accretion conformity. Consider the cartoon diagram shown in Figure~\ref{fig:cartoon2}. We show a primary and secondary halo pair embedded in a large-scale environment of tertiary halos. The top panel illustrates an under-dense, void-like environment with a scarcity of tertiaries; the bottom panel an over-dense environment crowded with many tertiaries.

In the over-dense environment, both primary and secondary halos tend, on average, to be close enough to some other tertiary halo such that $\rhillmin$ impinges on $\rvir.$ Conversely, in the under-dense environment, the typical $\rhillmin$ is substantially larger than $\rvir$ for both primaries and secondaries.  Coupling this observation with Fig.~\ref{fig:mar_vs_rhill}, we can intuitively see that evolving in a crowded environment tends to stifle present-day accretion rates, and conversely for under-dense environments. This effect will result in primary halos being surrounded by secondaries with correlated $\mar,$ the definition of halo accretion conformity.  

Note that in this diagram, {\em the tertiary halos producing $\rhillmin$ are distinct.} As we will see, the $\mar$ correlation between neighboring halos is {\em not} due a direct tidal force on the primary and secondary by the {\em same} massive tertiary, nor due to direct tidal interaction between the primary and secondary. We  justify this interpretation of the physical cause of halo accretion conformity in \S\ref{subsubsec:justification}. Before doing so, in the next section, \S\ref{subsubsec:lowzconform}, we quantify the scale- and mass-dependence of conformity in low-redshift halos. 

\begin{figure*}
\centering
\includegraphics[width=8cm]{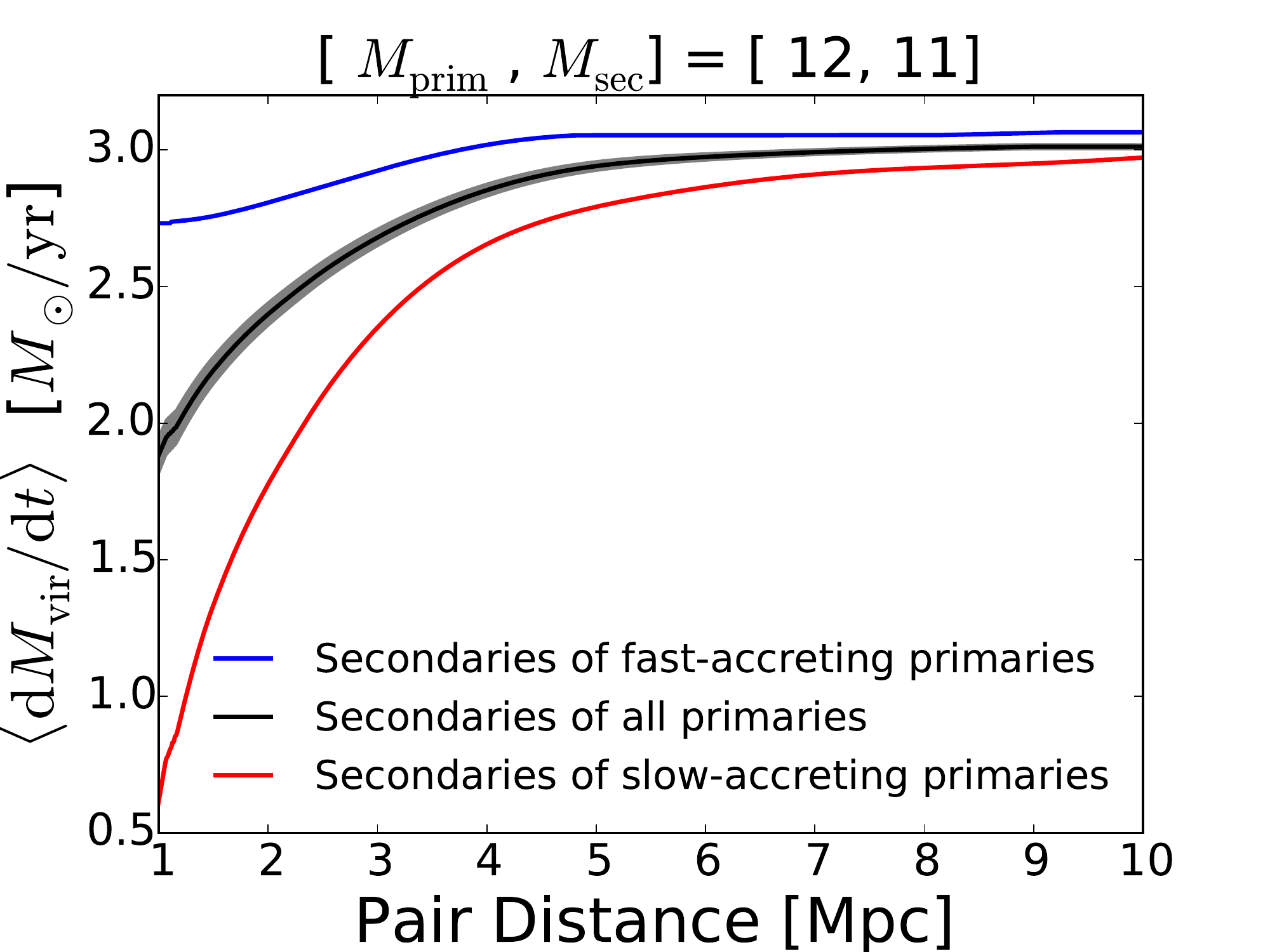}
\includegraphics[width=8cm]{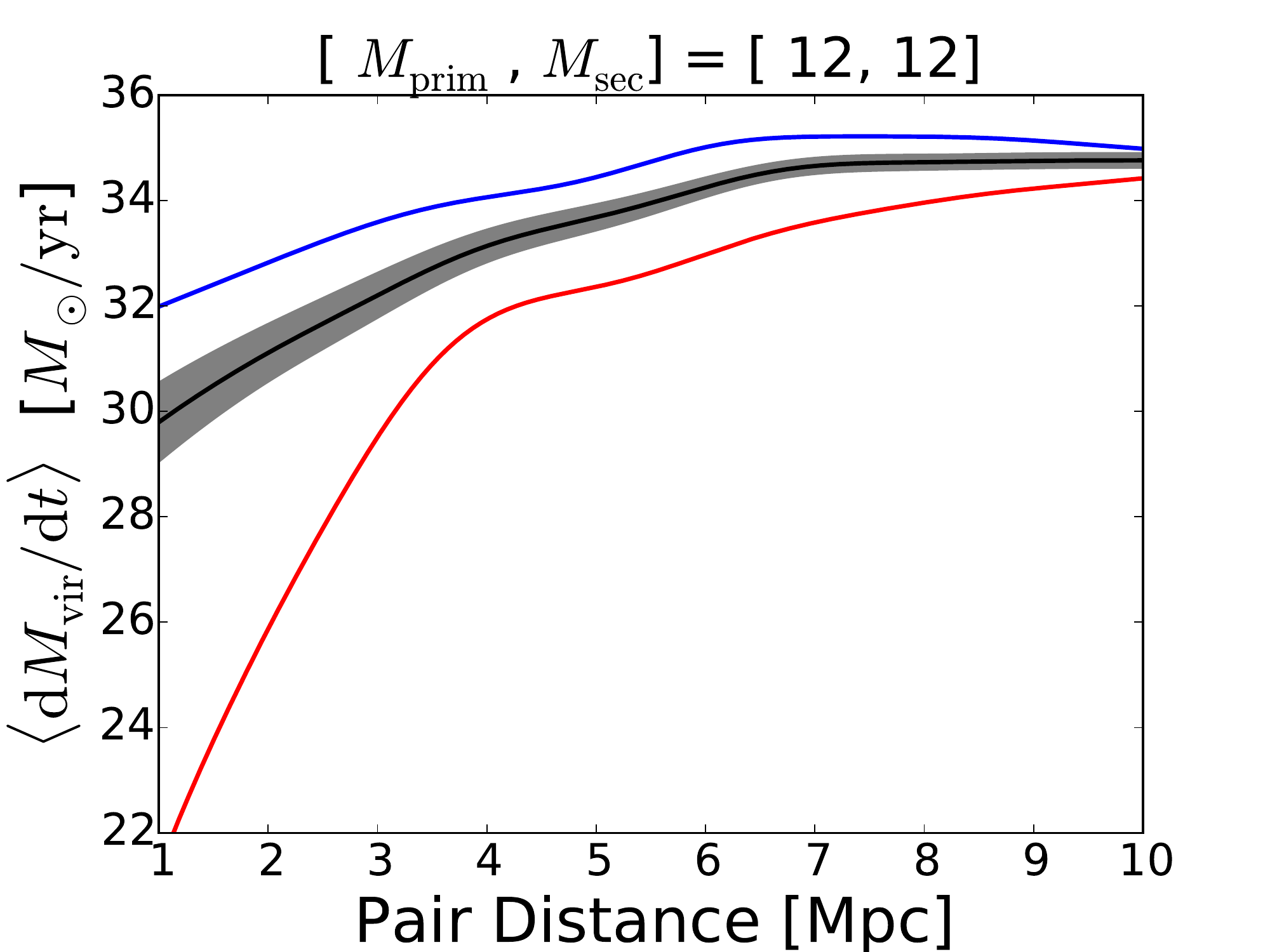}
\includegraphics[width=8cm]{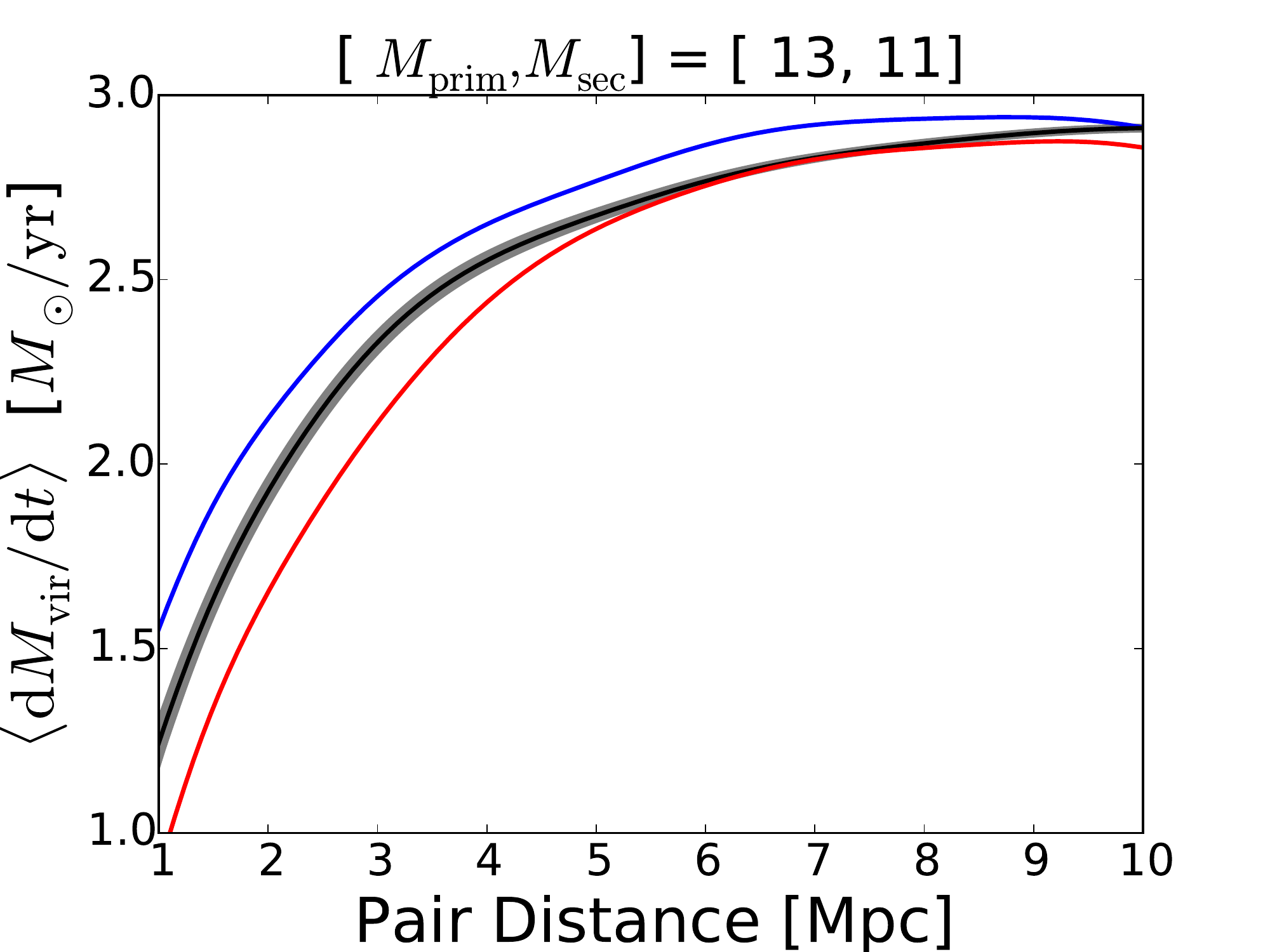}
\includegraphics[width=8cm]{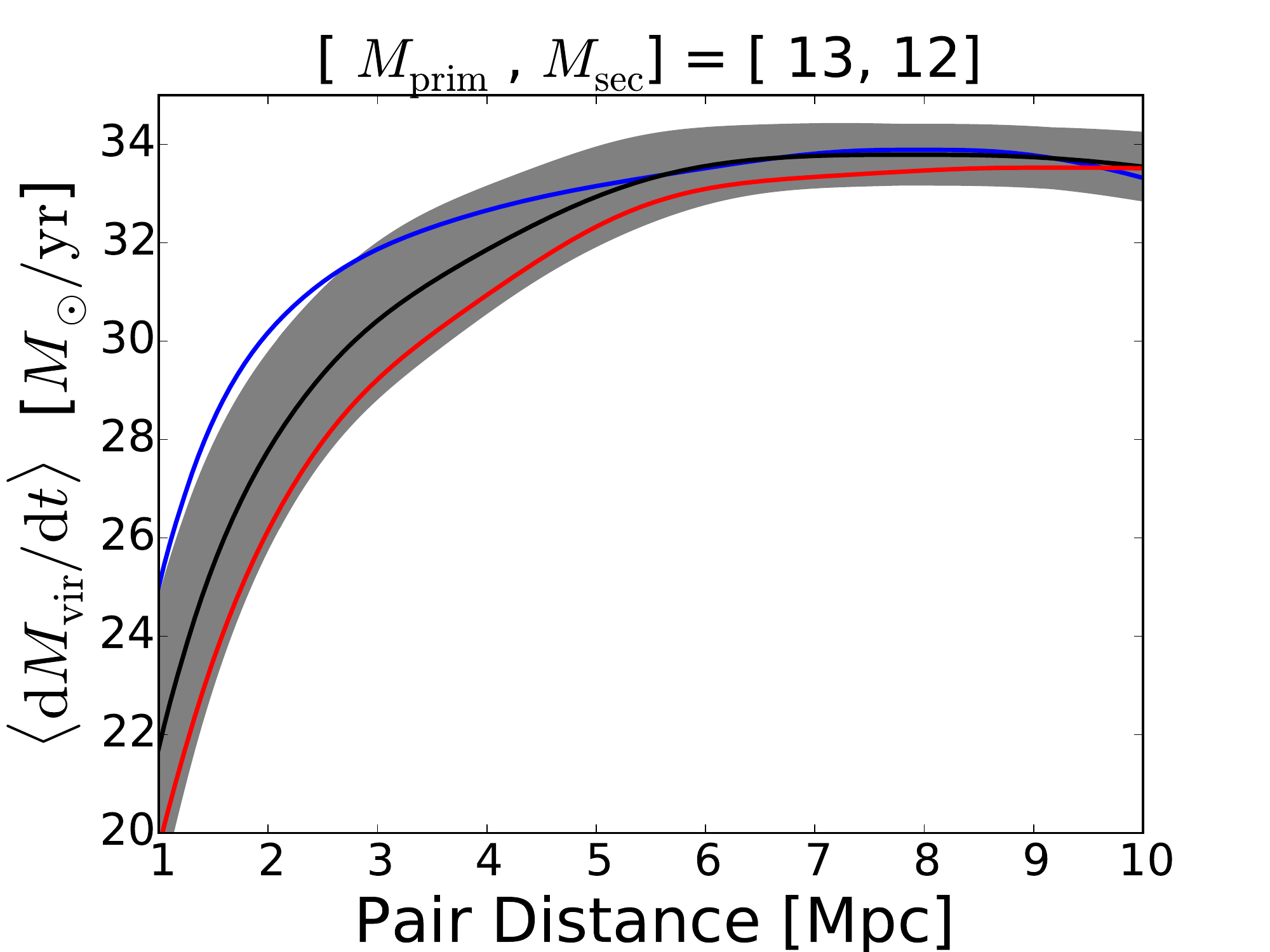}
\caption{{\bf Halo accretion conformity at $z=0.$} We compute $\dd\mvir^{\rm sec}/{\rm dt}$ of lower mass, secondary host halos in spherical shells surrounding higher mass, primary host halos. In each panel, the {\em black curve} shows the mean value of $\dd\mvir^{\rm sec}/{\rm dt}$ as a function of the distance between the primary and secondary. The masses of the primary/secondary pair are indicated by the title of each panel, where we have used mass-bins $0.35$ dex in total width; gray bands denote the jackknife error-on-the-mean.  Additionally, we have repeated each calculation after first sub-dividing the primary halos into quartiles of $\dd\mvir^{\rm prim}/{\rm dt}.$ The mean $\dd\mvir^{\rm sec}/{\rm dt}$ of secondary halos surrounding the fastest- and slowest-accreting quartile of primaries is shown with the {\em blue and red curves}, respectively. The upshot is that  {\em i}) fast-accreting primaries tend to be surrounded by fast-accreting secondary halos, and conversely; {\em ii})  both the mass- and scale-dependence of this signal closely resembles the conformity signal in SDSS observations of central galaxy SFR. 
}
\label{fig:upshot}
\end{figure*}

\begin{figure*}
\centering
\includegraphics[width=8.5cm]{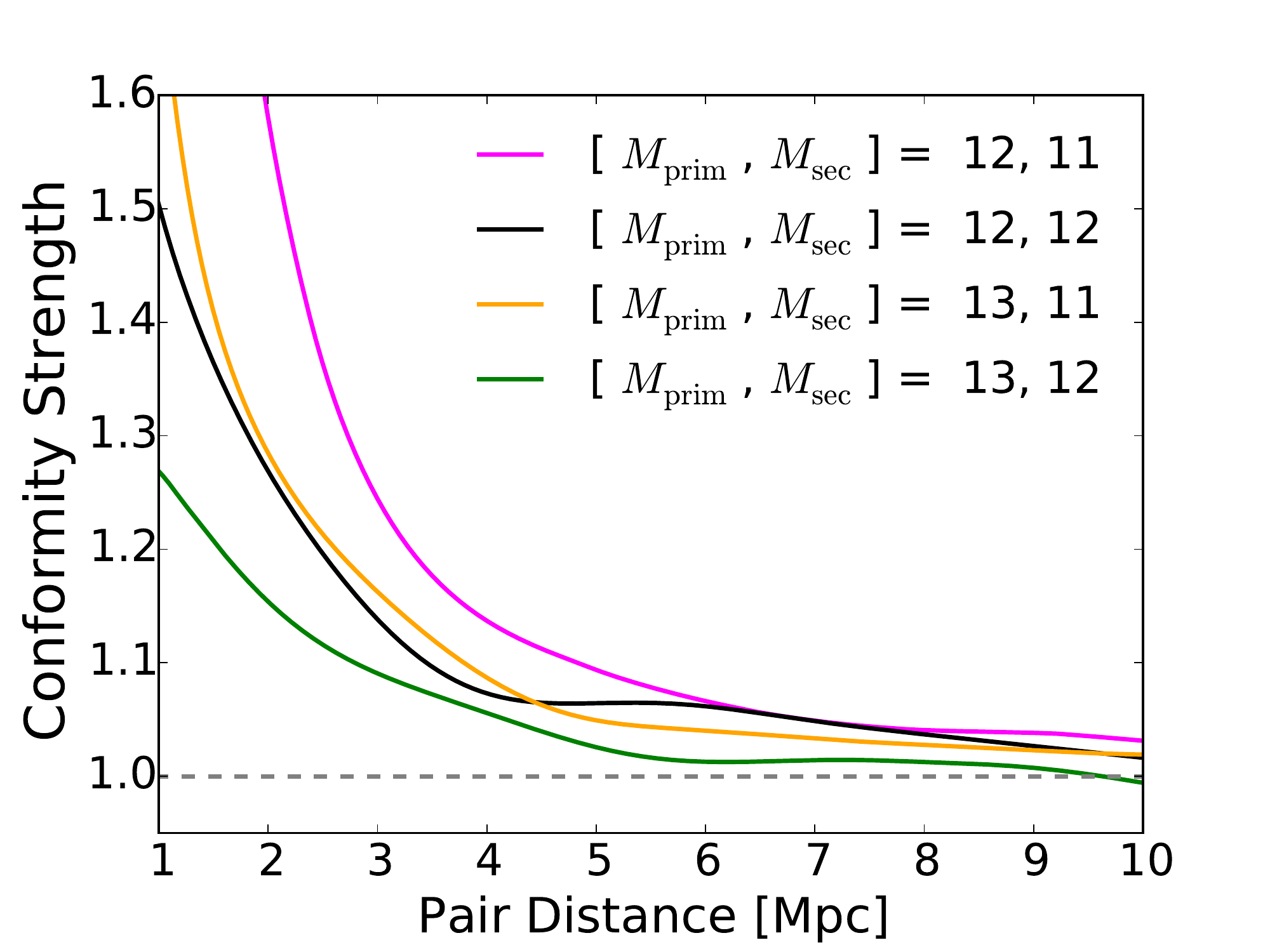}
\caption{{\bf $\mvir-$dependence of halo accretion conformity strength}. 
We plot the ratio of the blue-to-red curves in each of the panels in Fig.~\ref{fig:upshot}, referring to this ratio as the ``conformity strength.'' Halo accretion rates are correlated out to $R\sim10\mpc,$ a trend that weakens with increasing mass. 
}
\label{fig:upshotratio}
\end{figure*}


\subsubsection{Conformity at low-redshift}
\label{subsubsec:lowzconform}

In Figure~\ref{fig:upshot} we show the primary result of this work: the measurement of halo accretion conformity.  Roughly speaking, Figure~\ref{fig:upshot} shows the mass- and scale-dependence of how the mass accretion rates of lower-mass secondary halos ``know about" the mass accretion rates of higher-mass primary halos. We separately study this signal for several different choices of primary and secondary halo mass, as shown in the different panels. In detail, we calculated the results in Figure~\ref{fig:upshot} as follows. 

For each halo $P_{\rm i}$ in a given sample of primaries, we identify all secondary halos located in a spherical shell of radius $\dphys$ about the primary. We then calculate $\langle\mar\rangle_{P_{\rm i}}(\dphys),$ the mean $\mar$ of the secondaries in the shell that surrounds primary $P_{\rm i}.$ The vertical axis in all panels of Figure~\ref{fig:upshot} shows the mean of all these accretion rates, weighting each primary equally. We choose this method (rather than giving equal weight to each secondary) so that primaries in over- and under-dense environments are treated on equal footing. We note, though, that our results are qualitatively insensitive to this choice. 

The {\em black curves} in Figure~\ref{fig:upshot} show the results of the above calculation for all primary and secondary halos whose masses are indicated at the top of each panel. In direct analogy to the observational measurement of galactic conformity, we have repeated the above exercise after first sub-dividing the sample of primaries into quartiles of primary $\mar.$ In each panel, the {\em blue curve} shows  $\langle\mar\rangle$ of secondaries surrounding the fastest-accreting quartile of primaries; the {\em red curve} shows the same, but for secondaries surrounding the slowest-accreting quartile of primaries.

The most important feature of Figure~\ref{fig:upshot} is the separation of the red and blue curves. This indicates that fast-accreting primary halos tend to be surrounded by an environment of fast-accreting secondaries, and conversely for slow-accreting primaries. For lower mass halos, this separation persists to $10$ $\mpc,$ showing that {\em halo mass accretion rates are correlated for spatial separations greater than thirty times the virial radius.} 

In Figure~\ref{fig:upshotratio} we illustrate the mass-dependence of halo accretion conformity in more detail. The vertical axis in Figure~\ref{fig:upshotratio} shows the ratio of each blue/red curve in Figure~\ref{fig:upshot}---i.e., a measure of how secondary halo accretion rates increase nearby fast-accreting primaries.   Each curve in Figure~\ref{fig:upshotratio} corresponds to a particular panel in Figure~\ref{fig:upshot}, as indicated by the legend. Larger values of the vertical axis in Figure~\ref{fig:upshotratio} correspond to stronger halo accretion rate conformity. 

The mass dependence of conformity appears to be quite simple: {\em lower-mass halos exhibit stronger conformity}. This is to be expected based on the results in \citet{hahn_etal09}. Conformity is an effect caused by the large-scale environment, which has a proportionally greater effect on halos of smaller mass. We note that both the mass- and scale-dependence of the signal shown in Figures~\ref{fig:upshot} \&~\ref{fig:upshotratio} closely resembles the trends seen in the SSFRs and gas fractions of SDSS galaxies \citep{kauffmann_etal13}.

\subsubsection{Quantitative justification of physical picture}
\label{subsubsec:justification}

While the physical picture sketched in \S\ref{subsubsec:intuition} is intuitive, it is not obvious that the tidal forces typically arise from {\em different} tertiary halos.  Indeed, primary and secondary halos that are both close to the same group or cluster may have suppressed $\mar$ due to {\em direct} tidal interaction with the {\em same} massive halo.  This proposed explanation differs qualitatively from the one illustrated in Figure \ref{fig:cartoon2}; in this alternative scenario, massive groups and clusters play a privileged role in how they impact the assembly history of the low-mass halos in their $\sim5$ $\mpc$ environment. In this section we conclusively rule out this interpretation: {\em the halos of massive groups and clusters do not have special significance in the physics driving halo accretion conformity. }

\begin{figure}
\centering
\includegraphics[width=8cm]{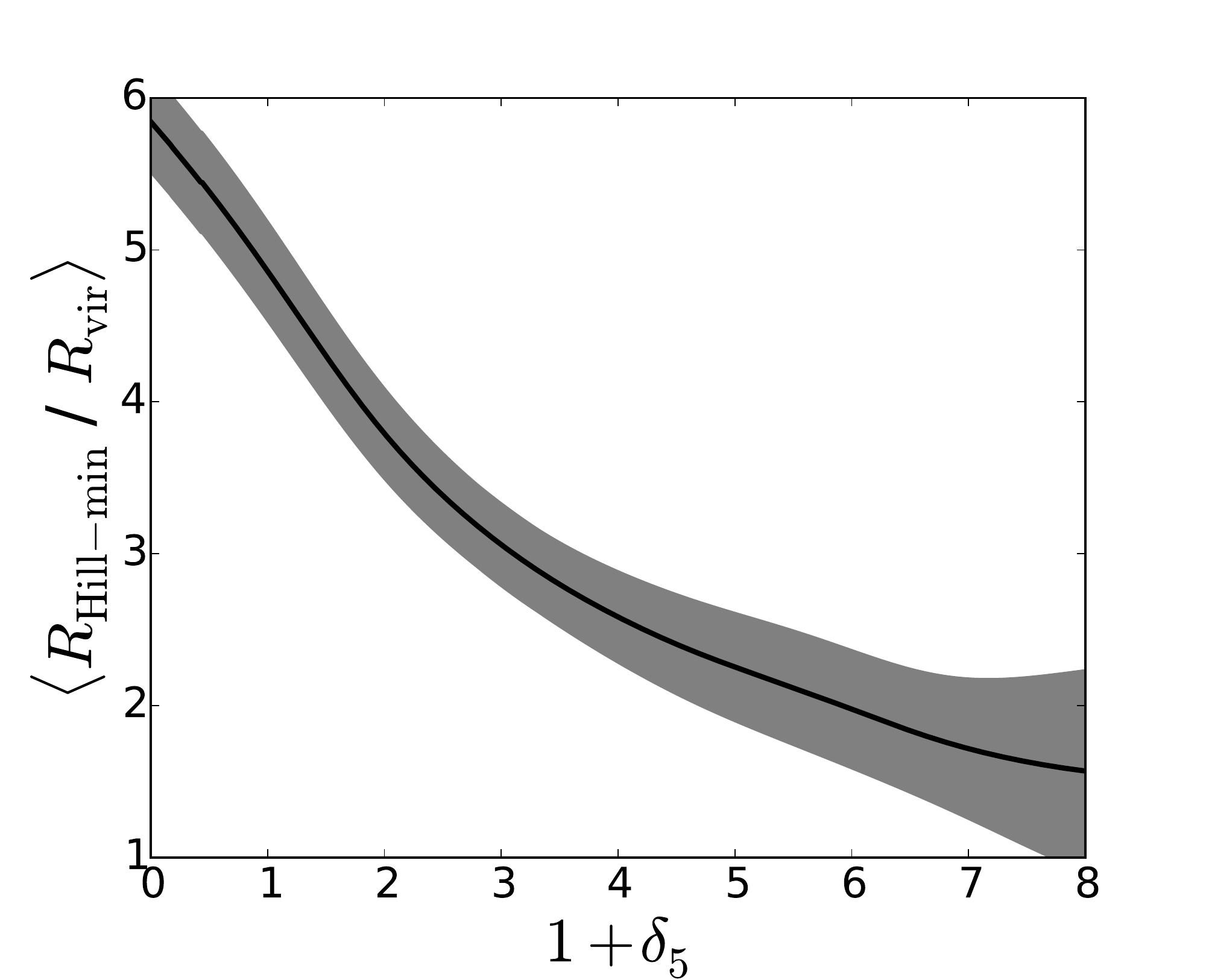}
\caption{{\bf Denser environments have stronger tidal fields.} For $\mvir=10^{11}\msun$ halos at $z=0,$ as a function of $5$-$\mpc$ over-density we show the mean tidal field strength, as quantified by $\rhillmin.$ 
}
\label{fig:rhill_density}
\end{figure}


\begin{figure}
\centering
\includegraphics[width=9cm]{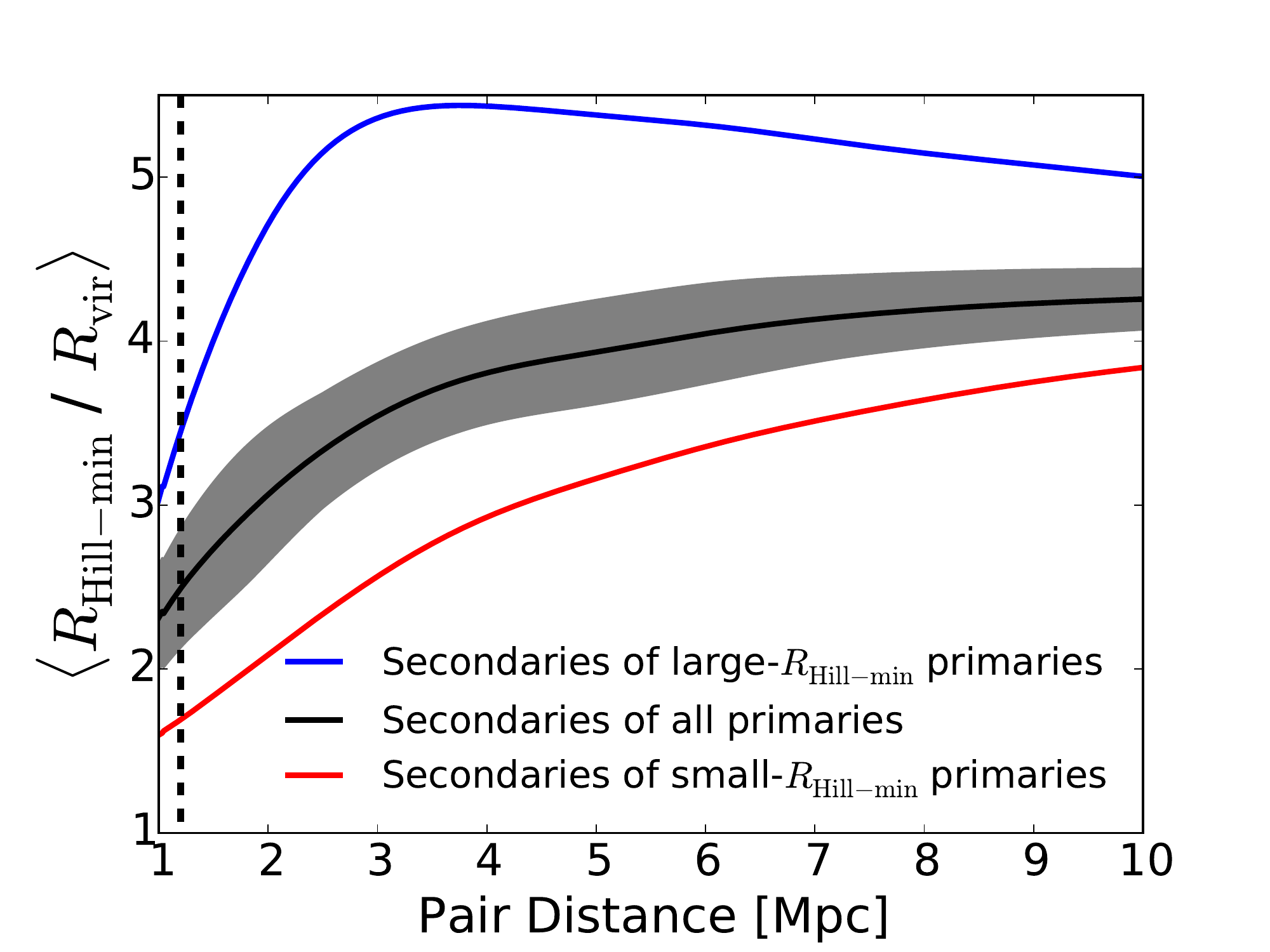}
\caption{{\bf Tidal fields are correlated on large scales}. We show the mean value of $R_{\rm hill-min}$ of secondary halos in spherical shells surrounding primaries. In analogy to Fig.~\ref{fig:upshot}, we have repeated the calculation after first dividing the primary sample into quartiles of $R^{\rm prim}_{\rm hill-min}.$ The mean $R_{\rm hill-min}$ of secondaries surrounding primaries in the top (bottom) quartiles of $R^{\rm prim}_{\rm hill-min}$ is shown with the blue (red) curve. With the dashed, vertical black line we show the distance at which $R_{\rm hill}=3\rvir$ (see Fig.~\ref{fig:mar_vs_rhill}).  {\em Primary halos in a strong tidal field are surrounded by secondaries that are also in a strong tidal field, the origin of halo accretion conformity.}
}
\label{fig:rhill_vs_dist}
\end{figure}


First, in Figure~\ref{fig:rhill_density} we show the strong correlation between large-scale environmental density and tidal field strength. For $z=0$ {\em Bolshoi} halos in a $0.35$ dex-width bin bracketing $\mvir=10^{11}\msun,$ we show the relationship between $\langle\rhillmin\rangle$ and environmental density. For our density estimator, we use $1+\delta_5,$ computed as the mass density in a $5$ $\mpc$ sphere divided by the cosmic mean matter density.  As expected, the denser the environment, the smaller the typical value of $\rhillmin,$ and the stronger the tidal field. 

Second, in Figure~\ref{fig:rhill_vs_dist} we show explicitly that a primary halo's tidal field is correlated with the tidal field of secondary halos out to large distances. We proceed in an exactly analogous fashion as we did to calculate the results shown in Figure~\ref{fig:upshot}. Here we focus on primary halos of mass $\mvir=10^{12}\msun,$ and secondary halos of mass $\mvir=10^{11}\msun.$ In spherical shells surrounding each primary, we compute the mean value of $\rhillmin$ of the secondaries in each shell, and then average the shells. Thus the {\em black curve} in Figure~\ref{fig:rhill_vs_dist} shows the average $\rhillmin$ value of the secondaries as a function of their distance to the primary. The closer a $10^{11}\msun$ secondary halo is to a $10^{12}\msun$ primary, the denser the environment of the secondary, and the stronger the tidal field exerted by the environment on the secondary.

Again we divide our sample of primaries into quartiles, this time based on the $\rhillmin$ value of the primary. The mean tidal field strength of secondaries surrounding primaries in the {\em weakest} tidal field are shown with the {\em blue curve}; results for secondaries surrounding primaries in the {\em strongest} quartile of tidal field strength are shown with the {\em red curve}. The separation of the blue and red curves in Figure~\ref{fig:rhill_vs_dist} tells us that primary halos in a strong tidal field tend to be surrounded by secondary halos also experiencing strong tidal forces, an intuitive result. 

With the vertical dashed line in Figure~\ref{fig:rhill_vs_dist}, we show the distance at which $\rhill=3 R_\mathrm{sec}.$ As shown in Figure~\ref{fig:mar_vs_rhill}, this is the tidal field strength necessary to have an appreciable impact on $\mar.$ Thus in order for the halo pair to exhibit a {\em direct} tidal influence on each other that is sufficiently strong to impact $\mar,$ they must be separated by a smaller distance than what is shown by the vertical dashed line. The separation between the red and blue curves remains large for separations that vastly exceed this distance. It is then not possible that halo accretion conformity is caused by direct tidal interaction between the halo pair. 

\begin{figure}
\centering
\includegraphics[width=9cm]{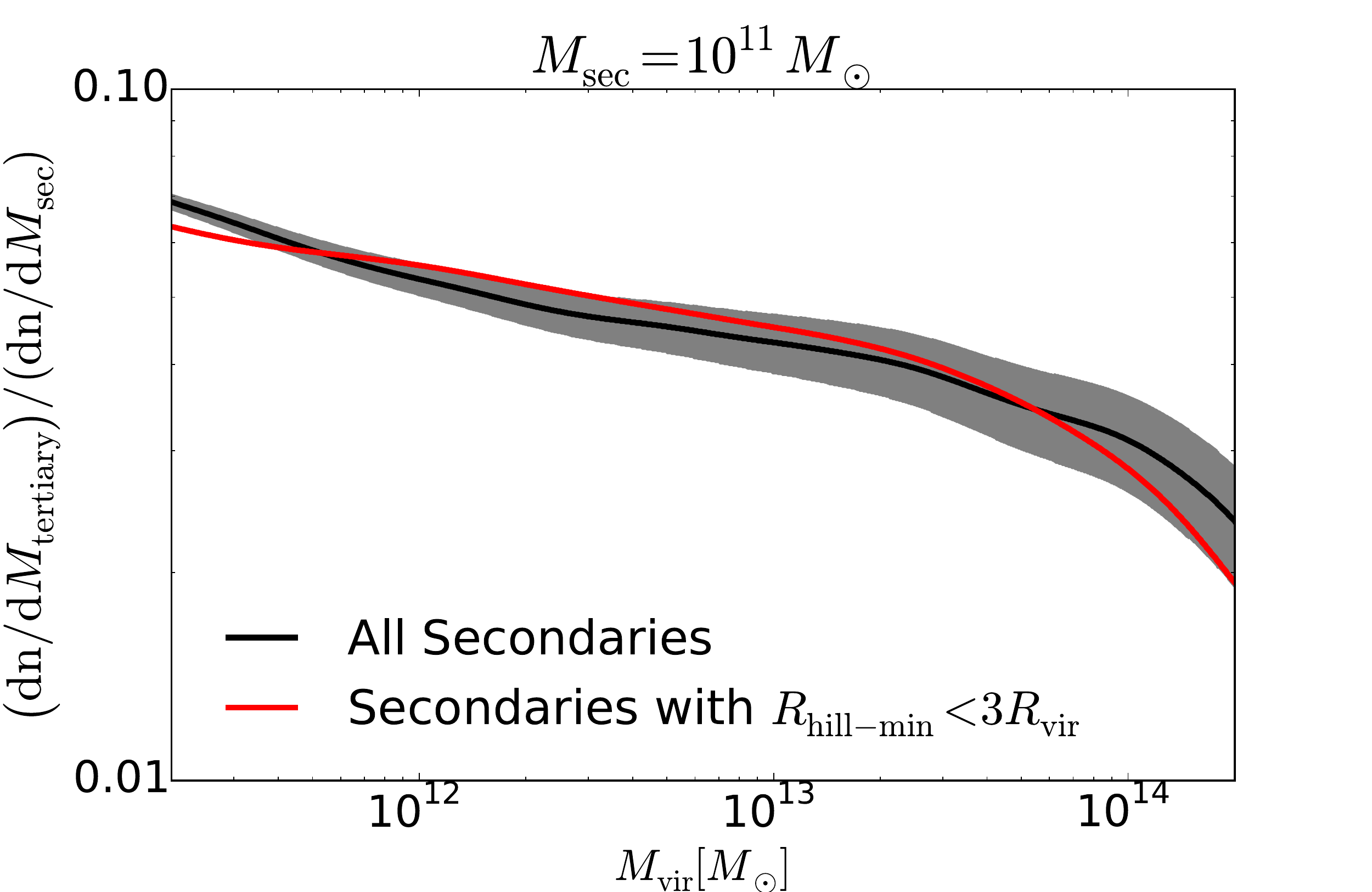}
\includegraphics[width=9cm]{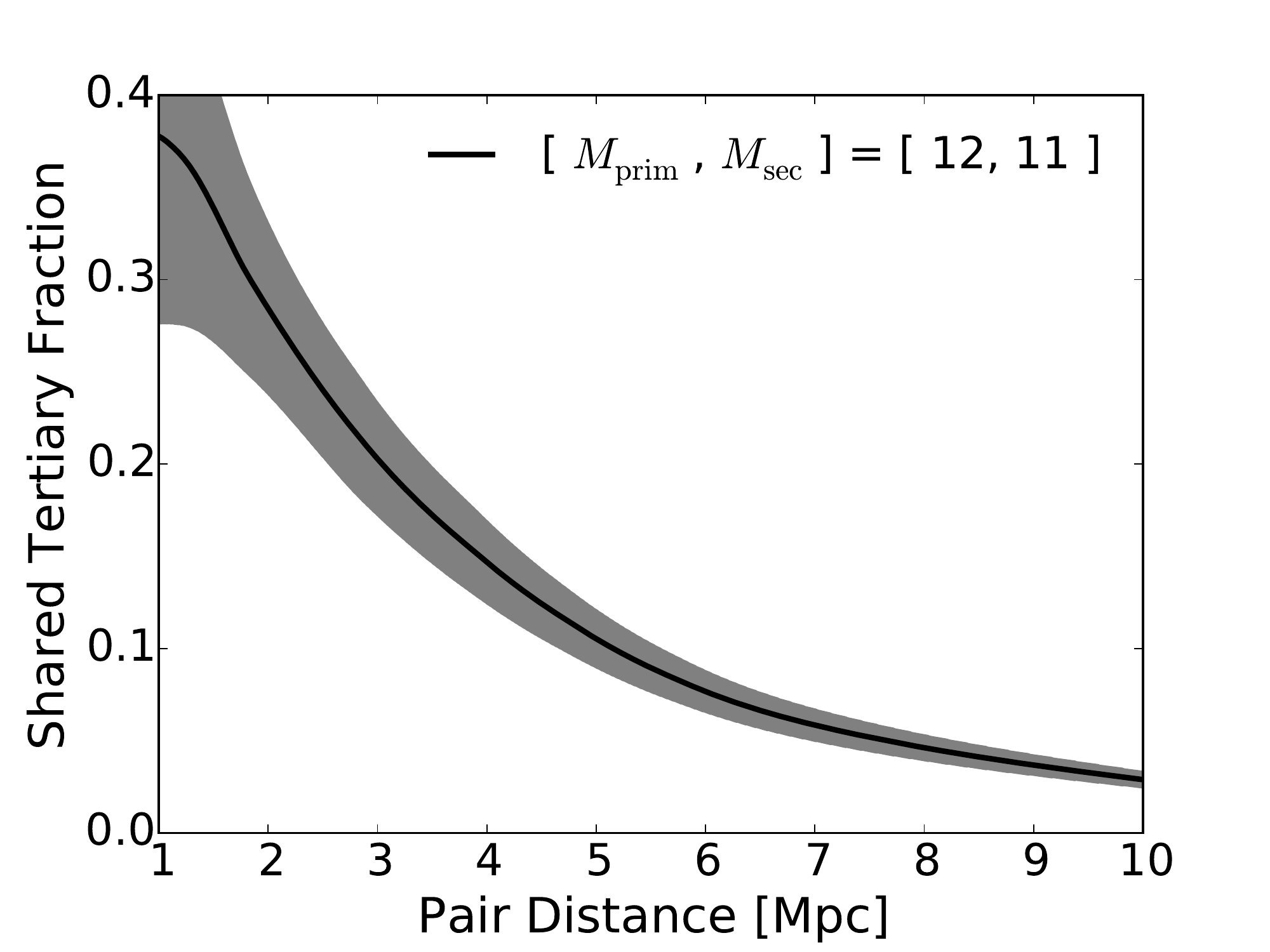}
\caption{{\bf Massive groups and clusters do not have special significance in the physics of halo accretion conformity.} 
{\em Top Panel:} For $\Msec=10^{11}\msun$ secondary halos, the {\em black curve} shows the mass function of the tertiary halos responsible for $\rhillmin,$ normalized by the mass function of the secondaries. The strikingly similar {\em red curve} shows the same quantity, but only for secondaries with $\rhillmin<3\rvir.$ {\em Bottom Panel:} We calculate the fraction of pairs with the {\em same} tertiary halo responsible for $\rhillmin$ as a function of the distance between the halo pair. Taken together, the two panels imply that halo accretion conformity is not preferentially influenced by massive groups/clusters, nor is the signal due to mutual proximity to the same object, justifying the picture illustrated in Fig.~\ref{fig:cartoon2}. 
}
\label{fig:sharedtertiary}
\end{figure}

Finally, in Figure~\ref{fig:sharedtertiary} we assess the plausibility of the alternative scenario described in the beginning of this section. Again we focus on secondary halos of mass $\Msec=10^{11}\msun,$ where the effects are largest. First, with the {\em black curve} in the {\em top panel}, we show the mass function of the tertiary halos that are responsible for the $\rhillmin$ of the secondaries; we have scaled this mass function by the abundance of the secondaries, so that the vertical axis in Figure~\ref{fig:sharedtertiary} is in dimensionless units. The {\em red curve} shows the same result, but only for secondaries with $\rhillmin<3\rvir;$ we remind the reader that Figure~\ref{fig:mar_vs_rhill} implies that this is the tidal field strength required to impact $\mar.$ {\em The similarity of the red and black curves in Figure~\ref{fig:sharedtertiary} implies that the $\mar-$stifling tidal forces experienced by the secondaries are not preferentially exerted by massive groups or clusters. }

In the {\em bottom panel}, we compute the fraction of primary/secondary halo pairs for which the same tertiary halo is responsible for the $\rhillmin$ of both the primary and secondary. Figure~\ref{fig:sharedtertiary} shows how this fraction varies as a function of the distance between the pair. At distances $\dphys\approx5$ $\mpc,$ where both the tidal field strength (Figure~\ref{fig:rhill_vs_dist}) and mass accretion rates (Figure~\ref{fig:upshot}) are still significantly correlated, over $90\%$ of primary-secondary halo pairs have distinct tertiary halos responsible for the most significant tidal field. This rules out the possibility that mutual proximity to the same massive object can explain correlations between low-mass halo accretion rates. The singular presence of a massive group or cluster has no special significance in the physics of halo accretion conformity. 

\subsection{Conformity at High Redshift}
\label{subsec:highzresults}

We conclude our results by showing how halo accretion conformity evolves with redshift. We repeat the calculation illustrated in Figures~\ref{fig:upshot} \& \ref{fig:upshotratio}, but now including {\em Bolshoi} halos at redshifts $z=0.5, 1$ and $2.$ We focus exclusively on a fixed pair of primary and secondary halo masses of mass $10^{12}\msun$ and $10^{11}\msun$, respectively, where the $z=0$ signal is strongest. These calculations allow us to estimate how the observed galactic conformity signal should evolve with redshift under the assumption that halo accretion rates are correlated with galaxy star formation rates (see \S\ref{subsec:highz} for further discussion).  

\begin{figure*}
\centering
\includegraphics[width=8cm]{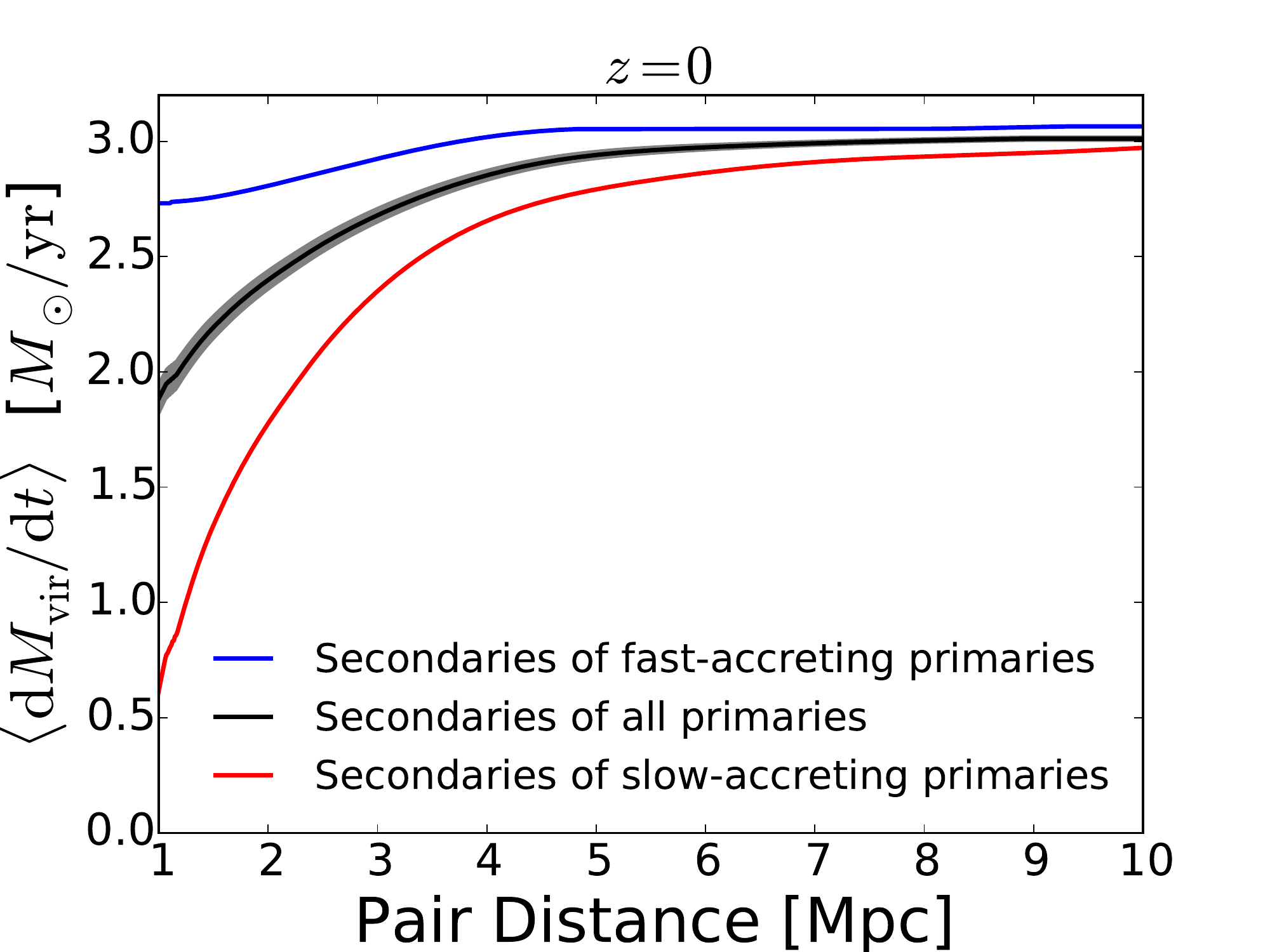}
\includegraphics[width=8cm]{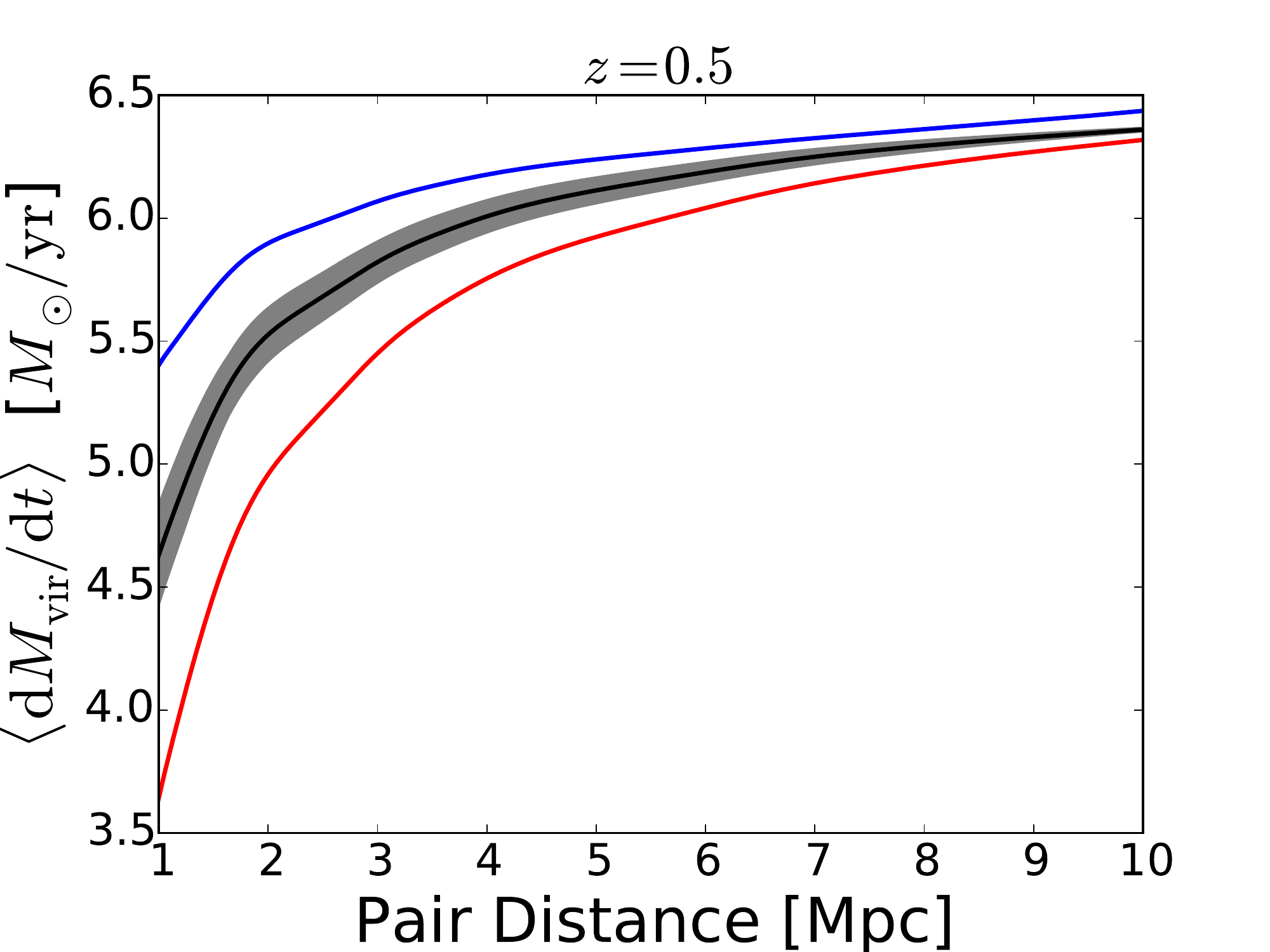}
\includegraphics[width=8cm]{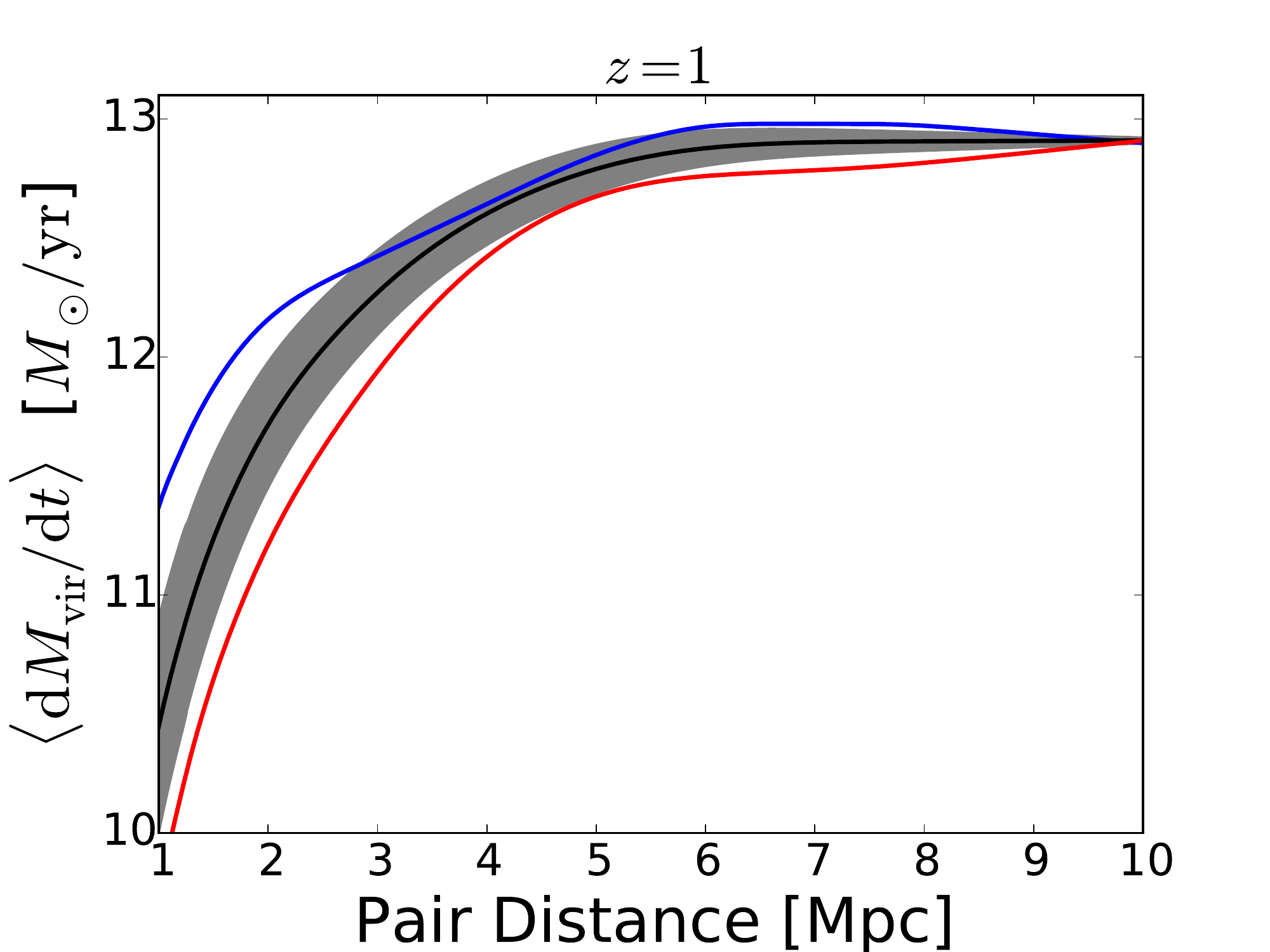}
\includegraphics[width=8cm]{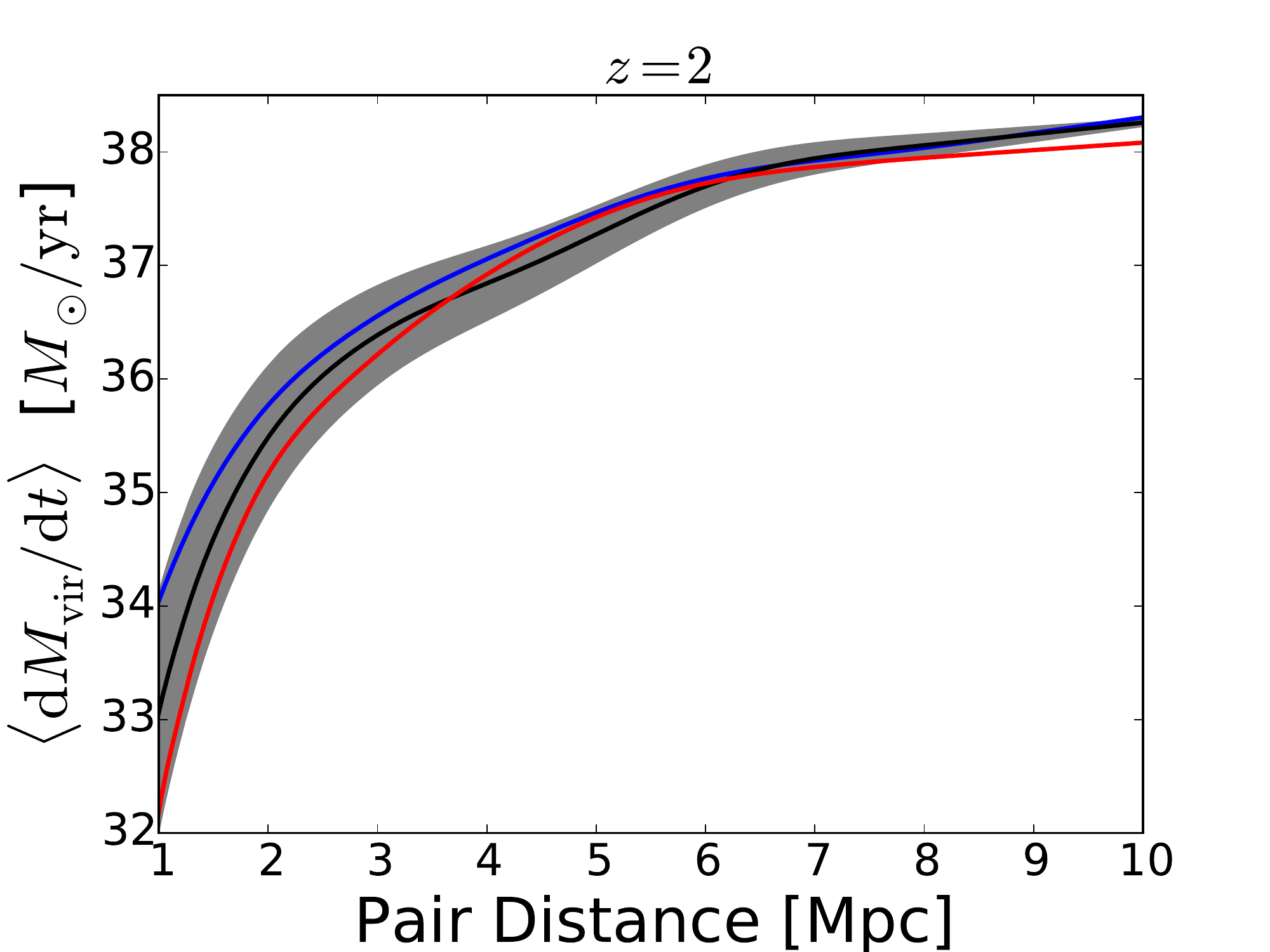}
\caption{ {\bf Halo accretion conformity at high-redshift.} Same as Fig.~\ref{fig:upshot}, but for halos at different redshifts, labeled by the title of each panel. All curves pertain to primaries of mass $\mvir=10^{12}\msun$ and secondaries with $\mvir=10^{11}\msun$ at the indicated redshift. Halo accretion conformity generically weakens at higher redshift, due primarily to the time evolution of collapse mass. 
}
\label{fig:zdep}
\end{figure*}

\begin{figure*}
\centering
\includegraphics[width=8.5cm]{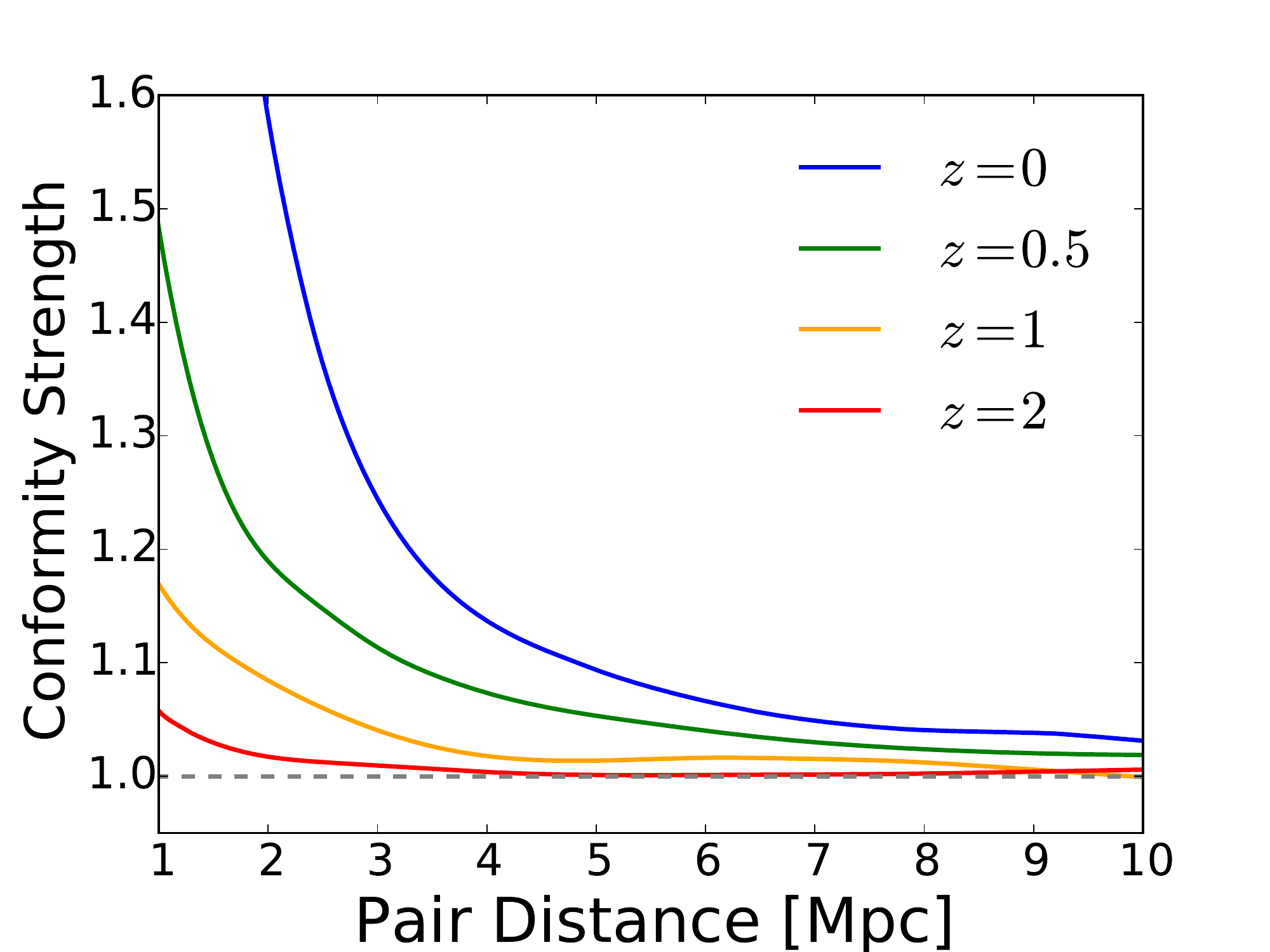}
\caption{ {\bf Redshift-dependence of halo accretion conformity strength}. Same as Fig.~\ref{fig:upshotratio}, but for halos at different redshifts. Each curve shown in this figure is computed as the ratio of the corresponding blue-to-red curves in Figure~\ref{fig:zdep}. For redshifts $z\gtrsim1,$ conformity on large scales ($R\gtrsim3$ Mpc) vanishes to undetectable levels.}
\label{fig:zdepstrength}
\end{figure*}

Figure~\ref{fig:zdep} shows the direct analog of Figure~\ref{fig:upshot}, only showing high-redshift results. The strength of the signal substantially weakens as we go farther back in cosmic time. This weakening is more easily seen in Figure~\ref{fig:zdepstrength}, which shows the high-redshift analog of Figure~\ref{fig:upshotratio}. 

This weakening is as expected. Since collapse mass $M_{\rm coll}$ decreases with increasing redshift, by comparing the conformity signal for halos of fixed mass, at higher redshift we are  studying halos with larger values of $\mvir/M_{\rm coll}.$ We have already seen in Figure~\ref{fig:upshotratio} that conformity weakens for higher mass halos, a fact also seen in SDSS observations \citep{kauffmann_etal13}. Moreover, it has already been shown in \citet{hahn_etal09} that $M_{\rm coll}$ sets the natural mass scale for how large-scale environment impacts halo assembly.  It is therefore completely natural that the conformity signal weakens at higher redshift for halos of the same mass.  As discussed in \S\ref{subsec:highz}, we fully expect the same to hold true for future measurements of galactic conformity at higher redshift, which would provide strong observational support for the natural hypothesis that central galaxy SFR and dark matter halo accretion rate are correlated.

\section{Discussion}
\label{sec:discussion}

In \S\ref{subsec:coevolution} we describe how conformity provides information about the coupling between galaxy and halo growth; we also sketch how future measurements of galactic conformity at both low- and high-redshift can be used to guide the modeling of feedback processes in hydrodynamical simulations and semi-analytic models. We outline the fundamental connection to assembly bias in \S\ref{subsec:assembias}. We describe how our work suggests a connection between large- and small-scale conformity in \S\ref{subsec:onehaloconformity}, and discuss alternative quantifications of our results in \S\ref{subsec:altquant}. We conclude in \S\ref{subsec:highz} by discussing predictions for galactic conformity at higher redshifts. 

\subsection{The Coupling of Halo and Galaxy Growth}
\label{subsec:coevolution}

In the same work presenting the discovery of large-scale galactic conformity in SDSS data \citep{kauffmann_etal13}, it was also shown that a current semi-analytical model (SAM) of galaxy formation \citep{guo_etal11} predicted only a very weak conformity signal. Since our results show that conformity exists in the accretion rates of dark matter halos, the failure of this prediction is curious. After all, the star-formation history of a central galaxy in a SAM traces the mass assembly history of its host halo. 

To understand this puzzling failed prediction, it is useful to look to a model which {\em does} predict strong levels of galactic conformity: age matching  \citep{HW13a,hearin_etal13b,watson_etal14}. In the age matching model, older (earlier-forming) halos host central galaxies with older stellar populations relative to younger halos of the same mass. This naturally results in galactic conformity: the present work shows that fast-accreting (slow-accreting) halos tend to congregate together, and halo ages and accretion rates are tightly anti-correlated  at fixed mass \citep{hahn_etal09}. 

Since the \citet{HW13a} model presumes that halo and galaxy age are in monotonic correspondence at fixed stellar mass, with no scatter, this model represents {\em the limit of maximal coupling} between the growth of a galaxy and its parent halo. In the opposite extreme limit, empirical models such as the Halo Occupation Distribution (HOD) and Conditional Luminosity Function (CLF) represent {\em the limit of zero coupling,} since in these models present-day virial mass alone governs the properties of the halo's resident galaxies. HOD models, or indeed any model with zero coupling between central galaxy and halo growth, predict essentially zero two-halo conformity \citep{hearin_etal14,paranjape_etal15}. 

Thus, the coupling strength between halo assembly history and galaxy star formation directly influences the predicted conformity effect.  SAMs such as \citet{guo_etal11} heavily rely on present-day halo mass (or gravitational potential) to determine star formation rates, as do many other current semi-analytical models \citep[see][for a review]{YuLu14b}.  The presence of history-unaware feedback will of course tend to erase the memory the galaxy has of its parent halo's assembly, thereby erasing the conformity signal present in the underlying dark matter halo. 

We note that \textit{re}accretion of ejected gas can also play an important role for galaxy formation at $z=0$ \citep{Nelson15}.  If reaccretion were dominant, the same physical mechanism---i.e., external tidal forces---would still cause galactic conformity because the tidal forces would affect gas reinfall times.  In this case, the strength of conformity would also depend on the turnaround radius of ejected gas.  If the gas thermalized immediately or cycled nearby the galaxy, then external tidal forces would not play an important role, and no conformity would be present for the gas reinfall rate.  However, if galactic fountains propelled gas to significant fractions of the virial radius, then stronger external tidal forces would result in gas taking significantly longer to reinfall, reducing the total reinfall rate.  Hence, in this scenario, conformity would be connected to the strength of star formation feedback.

In either scenario, conformity represents a practically ideal statistic at $z=0$ to constrain the coupling strength between galaxy environment and galaxy growth, as well as the channels modulating the strength of feedback. Observational conformity measurements across cosmic time will provide invaluable insight into the physical processes regulating star formation and quiescence.

\subsection{The Fundamental Link between Halo Accretion Conformity and Assembly Bias}
\label{subsec:assembias}

Halo mass is the dominant variable determining the spatial distribution of dark matter halos. However, as discussed in \S\ref{sec:intro}, the clustering of halos has an additional dependence on formation time $z_{\rm form},$ a phenomenon dubbed {\em assembly bias}. This dependence can be quantified in terms of the relative strength of the two-point function of old and young halos of the same mass, as in \citet{gao_etal05}. 

This classical quantification of assembly bias is simply linked to conformity.  At fixed $\mvir,$ earlier forming halos have lower $\mar$ at $z=0.$ Thus since early-forming halos cluster more strongly than late-forming halos of the same $\mvir,$ then so do slow-accreting halos cluster more strongly than fast-accreting halos of the same $\mvir.$ Strongly clustered halos reside in preferentially denser environments, where tidal fields are stronger (Figure~\ref{fig:rhill_density}); the converse is true for weakly clustered halos. The basic result of this paper is that halos residing in the same tidal environment have correlated $\mar,$ and so we conclude that not only are conformity and assembly bias connected, {\em they are alternative statistical quantifications of the exact same phenomenon}. 

\subsection{Relevance to 1-Halo Conformity}
\label{subsec:onehaloconformity}

A correlation between the star-formation histories of neighboring galaxies was first discovered in \citet{weinmann06b}, who found that the SFR of satellite galaxies are correlated with the SFR of the associated central. By definition, central and satellite galaxies occupy the same dark matter halo, whereas the \citet{kauffmann_etal13} measurements pertain to galaxies occupying distinct halos. Accordingly, \citet{hearin_etal14} dubbed these two signals ``1-halo" and ``2-halo" conformity, respectively. 

This paper is chiefly concerned with the large-scale 2-halo conformity signal measured in \citet{kauffmann_etal13}. However, in light of the findings presented here, it is quite plausible that 1-halo and 2-halo conformity measurements probe the same underlying physics. After all, satellites lead most of their lives as centrals in the neighborhood of their ultimate host halo. Thus centrals living in host halos with below-average $\mar$ will tend to accrete satellites that, prior to infall, have themselves already experienced below-average $\mar,$ and conversely.  In fact, it is precisely this phenomenon which leads to the age matching model simultaneously exhibiting both 1-halo and 2-halo conformity (Campbell et al., in prep). In this way, 2-halo can naturally evolve into 1-halo conformity. 

\subsection{Alternative Quantifications}
\label{subsec:altquant}

Throughout this work, we have focused exclusively on $\rhillmin$ as the single number that encapsulates the influence of a halo's large-scale environment on its mass assembly. In developing our results, we explored a broad range of alternatives that could act as regulators of mass accretion. Our investigation of these alternatives included 
\ben
\item distance to the nearest halo of (variable) mass $\mvir;$
\item large-scale density (with a variable kernel);
\item host halo concentration, $V_{\rm max},$ and $z_{\rm form}$; 
\item whether or not the host halo has been previously ejected from another halo. 
\een
Of these alternatives, we find that $\rhillmin$ is the strongest predictor of halo accretion rate, regardless of the summary statistic of the $\mar$ distribution (median, mode, size of the low-$\mar$ tail, etc.). In addition to the clear physical motivation for $\rhillmin,$ this proxy  has the distinct advantage of being computable in a galaxy catalog, at least in principle. Redshift-space distortions will introduce noise into using $\rhillmin$ as an environmental proxy, but otherwise the only required ingredient is a model for the $M_{\ast}-\mvir$ relation. Existing models \citep[e.g,][]{behroozi13b,kravtsov13} may be sufficient for this purpose, since the mass-dependence of $\rhillmin$ is weak (see Eq.~\ref{eq:rhilldef}). 

We have also explored the dependence of halo accretion conformity on the timescale $\tau$ over which $\mar$ is defined (see Eq.~\ref{eq:taudyndef}). We find that the signal decreases in strength for $\tau\lesssim1$ Gyr, and essentially vanishes if  the ``instantaneous", snapshot-to-snapshot timescale is used. Since conformity in galaxy SFR is realized in the real Universe, this observation could signal a natural timescale over which galaxy and halo growth are correlated, or it could simply be a time- and/or mass-resolution effect. We leave this investigation as a subject for future work.

\subsection{Predictions for Conformity at Higher Redshifts}
\label{subsec:highz}

The qualitative redshift-dependence of conformity in Figs.\ \ref{fig:zdep} and \ref{fig:zdepstrength} translates straightforwardly to galaxies: {\em 2-halo galactic conformity should be much weaker at higher redshifts}. The logic of this expectation is quite simple. First, the stellar-to-halo mass relation evolves only very weakly with redshift \citep[e.g.,][]{behroozi13}. Thus by studying the redshift evolution of conformity at fixed $\mvir,$ we are effectively studying the redshift-evolution at fixed stellar mass. Second, the intrinsic mass scale in the problem is $M_{\rm coll},$ the halo model collapse mass, which decreases monotonically with increasing redshift, so that $$\frac{\mvir}{M_{\rm coll}(z>0)} > \frac{\mvir}{M_{\rm coll}(z=0)}.$$  Finally, the influence of environment on a halo generically decreases as $\mvir/M_{\rm coll}$ increases. Therefore, {\em if the SSFR of a central galaxy is correlated with its host halo accretion rate, the prediction that conformity weakens at higher redshift follows inexorably.} 

The largest halo accretion conformity signal in this study occurs for $10^{12}\msun$ primary halos with $10^{11}\msun$ secondary halos.\footnote{We note that we expect halo accretion conformity to continue to increase in strength towards lower masses, but this remains to be seen with a higher resolution simulation.} These correspond to $\sim 10^{10}-10^{10.5}\msun$ primary galaxies with $10^{9.5}-10^{10}\msun$ secondary neighbor galaxies (roughly independently of redshift for $z<4$; \citealt{behroozi13}). Galaxies of these masses are currently probed with precise redshifts out to $z\sim 0.5$ in, e.g., the PRIMUS \citep{Moustakas13}, VIPERS \citep{garilli_etal14}, and DEEP2 \citep{Bundy06} surveys. Hence our high-redshift predictions can be tested even with present datasets. 

If the strength of galactic conformity is correlated with the strength of halo accretion conformity, then Fig.\ \ref{fig:zdepstrength} provides a basic guide for how to relate $z=0$ conformity signals to $z>0$ conformity signals. For example, Fig.\ \ref{fig:zdepstrength} indicates that the conformity strength observed at a separation of 4Mpc at z=0 should be matched by a signal of the same strength at a separation of $\sim$2.2 Mpc (comoving) at $z=0.5$.  The following fit well-approximates the expectation from Fig.\ \ref{fig:zdepstrength}:
\begin{eqnarray}
\label{eq:highzprediction}
S_g(R, z) & \approx & 1 + A(z)\left(R + 1.65\right)^{B(z)} \\
A(z) & = & 15.8 - 6.5z \\
B(z) & = & -2.8 - 0.67z
\end{eqnarray}
where $R$ is the comoving separation between primary and secondary, $z$ is the redshift of the observation, and $S_g$ is the strength of galactic conformity, measured in the same fashion in Figures \ref{fig:upshotratio} \& \ref{fig:zdepstrength}, i.e., the upper-to-lower quartile ratio. 

We note that the prediction given in Eq.~\ref{eq:highzprediction} is only approximate, particularly because we have only studied conformity in three dimensional shells, rather than $2+1$-dimensional cylinders in redshift-space. Redshift-space distortions will generically reduce the signal, so the overall magnitude of the signal shown in Figures \ref{fig:upshotratio} \& \ref{fig:zdepstrength} represents an upper bound. However, we have not incorporated this effect into Eq.\ \ref{eq:highzprediction}, as the appropriate redshift window depends on the survey used.  Since the present literature does not even contain an order-of-magnitude estimate on how conformity should scale with redshift, we consider Eq.~\ref{eq:highzprediction} to be a useful guideline until more detailed models and measurements become available. 

For primary galaxies with stellar mass larger than $10^{11}\msun$ (hosted by $\sim 10^{13}\msun$ halos), Fig.~\ref{fig:upshotratio} suggests that the conformity signal is roughly half of that for the lower mass ($10^{10}-10^{10.5}\msun$) primary galaxies considered above.  This would suggest that the 2-halo galactic conformity signal would be marginally detectable at $z=0.5$, and only at closer distances to the primary galaxies ($\sim$1 Mpc).  However, the virial radius for a $10^{13}\msun$ host halo is already $\sim 0.6$ Mpc, and because galaxies continuously pass in and out of the virial radius, large halos can directly affect quenched fractions at several times their virial radii \citep{wetzel_etal13}.  Hence, it is likely that any signal found at higher redshift would be hard to attribute solely to 2-halo galactic conformity \citep[see also][for elaboration of this point]{paranjape_etal15}. 
Finally, we note that if 1-halo and 2-halo conformity are indeed manifestations of the same underlying physics, as discussed in \S\ref{subsec:onehaloconformity}, a corollary to our findings is the prediction that 1-halo conformity should also vanish for $z\gtrsim1.$

\section{Conclusions}

\label{sec:conclusions}

Our primary findings are:

\ben
\item The mass accretion rates of dark matter halos are highly correlated on large scales, as shown in Figures~\ref{fig:upshot} \& \ref{fig:upshotratio}. We dub this phenomenon {\em halo accretion conformity.} Both the scale- and mass-dependence are qualitatively similar to SDSS measurements of galactic conformity, strongly suggesting that at fixed mass, dark matter halo accretion rate and central galaxy SFR are correlated. 
\item The origin of halo accretion conformity is the mutual evolution of halos in the same large-scale tidal field (see Figures~\ref{fig:rhill_density} \& \ref{fig:rhill_vs_dist}). Massive groups and clusters play no special role in the physics giving rise to the signal (Fig.~\ref{fig:sharedtertiary}). 
\item We make quantitative predictions for the time evolution of galactic conformity (Eq.~\ref{eq:highzprediction}), suggesting that the signal will drop rapidly at higher redshifts. Our predictions are testable with existing datasets to $z\sim1.$ Future conformity measurements will be highly informative about the physical processes driving galaxy evolution. 
\een


\section{acknowledgments}

We thank Neal Dalal, Doug Watson and Joel Primack for useful discussions. We are particularly grateful to Shy Genel for physically insightful input in the formative stages of development. A portion of this work was also supported by the National Science Foundation under grant PHYS-1066293 and the hospitality of the Aspen Center for Physics. 
APH thanks The Cramps for their cover of {\em I Ain't Nothin' But a Gorehound.}  PSB was supported by a Giacconi Fellowship through the Space Telescope Science Institute, which is operated by the Association of Universities for Research in Astronomy, Incorporated, under NASA contract NAS5-26555. 

\bibliography{./marcorr}

\end{document}